\documentclass[11pt]{article}
\usepackage[dvipdfmx]{color}
\usepackage{graphicx}
\usepackage{xcolor}
  \usepackage{amsthm,amsfonts}
  \usepackage{amsmath}
\usepackage{mathabx}
\usepackage{slashed}
\usepackage{ulem}
\usepackage{cite}

\newcommand{\bea}   {\begin{eqnarray}}
\newcommand{\eea}   {\end{eqnarray}}

\begin{document}
\renewcommand{\thefootnote}{\fnsymbol{footnote}}

\thispagestyle{empty}

\title{${\mathbb Z}_2\times {\mathbb Z}_2$-graded parastatistics in \\multiparticle quantum Hamiltonians}
\author{ Francesco Toppan\thanks{{E-mail: {\it toppan@cbpf.br}}}
\\
\\
}
\maketitle

{\centerline{
{\it CBPF, Rua Dr. Xavier Sigaud 150, Urca,}}\centerline{\it{
cep 22290-180, Rio de Janeiro (RJ), Brazil.}}
~\\
\maketitle
\begin{abstract}
The recent surge of interest in ${\mathbb Z}_2\times {\mathbb Z}_2$-graded invariant mechanics poses the challenge of understanding the physical consequences of a ${\mathbb Z}_2\times{\mathbb Z}_2$-graded symmetry. \par In this paper it is shown that non-trivial physics can be detected in the multiparticle sector of a theory, being induced by the ${\mathbb Z}_2\times{\mathbb Z}_2$-graded parastatistics obeyed by the particles. \par
The toy model of the ${\cal N}=4$ supersymmetric/ ${\mathbb Z}_2\times {\mathbb Z}_2$-graded oscillator is used.
In this set-up the one-particle energy levels and their degenerations  are the same for both supersymmetric and ${\mathbb Z}_2\times{\mathbb Z}_2$-graded versions. Nevertheless, in the multiparticle sector, a measurement of an observable operator on suitable states can discriminate whether the
system under consideration is composed by ordinary bosons/fermions or by ${\mathbb Z}_2\times {\mathbb Z}_2$-graded particles. Therefore, ${\mathbb Z}_2\times {\mathbb Z}_2$-graded mechanics has experimentally testable consequences.  \par
Furthermore, the ${\mathbb Z}_2\times {\mathbb Z}_2$-grading constrains the observables to obey a superselection rule.\par
As a technical tool, the multiparticle sector is encoded in the coproduct of a Hopf algebra defined on a Universal Enveloping Algebra of a graded Lie superalgebra with a braided tensor product.

\end{abstract}
\vfill
\rightline{CBPF-NF-007/20}
\newpage

\section{Introduction}
This paper presents a theoretical test of the physical consequences of the ${\mathbb Z}_2\times{\mathbb Z}_2$-graded parastatistics in a toy model case of a quantum Hamiltonian. This work is at a crossroad of two independent,
but related, lines of research which are both based on applications of the ${\mathbb Z}_2\times{\mathbb Z}_2$-graded Lie superalgebras first introduced in \cite{{rw1},{rw2}}. The first line, which consists of physical models possessing a ${\mathbb Z}_2\times{\mathbb Z}_2$-graded symmetry, has recently received a considerable attention (see \cite{{akttll1},{akttll2},{brdu},{aktclass},{aktquant},{brusigma}}). The second line, concerning parastatistics induced by ${\mathbb Z}_2\times {\mathbb Z}_2$-graded superalgebras, has been investigated in \cite{{yaji1},{yaji2},{kd1},{kaha},{kaha2},{kan},{tol2},{stvdj}}.\par
The motivation of the paper stems from an open question which finds here its answer. The first derived models of  quantum mechanics with 
${\mathbb Z}_2\times{\mathbb Z}_2$-graded one-dimensional Poincar\'e invariance, see \cite{{brdu},{aktquant}}, are also examples 
of Supersymmetric Quantum Mechanics. Then, the natural question which emerges is whether the ${\mathbb Z}_2\times{\mathbb Z}_2$-graded symmetry is redundant (a nice further structure to describe these models, but void of physical consequences not already encoded in supersymmetry) or whether it has measurable effects.  For the class of  single-particle quantum Hamiltonians presented in both \cite{brdu} and \cite{aktquant} the  ${\mathbb Z}_2\times{\mathbb Z}_2$-graded symmetry
is redundant. We will show in this article that this is no longer true for the class of multiparticle  quantum Hamiltonians introduced in \cite{aktquant}. In this way the  ${\mathbb Z}_2\times{\mathbb Z}_2$-graded parastatistics comes into play: the consistent (anti)symmetrizations of the wave functions produce testable differences if the particles are assumed to be
ordinary bosons/fermions (in the case of supersymmetry) or ${\mathbb Z}_2\times{\mathbb Z}_2$-graded particles.\par
Before delving more into the results of this paper, let us present at first the context, that is ${\mathbb Z}_2\times{\mathbb Z}_2$-graded Lie superalgebras and their applications both as symmetries of physical models and as a framework for a class of parastatistics.\par
The ${\mathbb Z}_2\times{\mathbb Z}_2$-graded Lie superalgebras were introduced in \cite{{rw1},{rw2}} (even if some related structures were already investigated in \cite{Ree}). These works extended the notion of ordinary, ${\mathbb Z}_2$-graded, Lie superalgebras appearing in \cite{kac} and suggested possible applications to elementary particles. Ever since the mathematical aspects (including classifications, representations, etc.) of the ${\mathbb Z}_2\times{\mathbb Z}_2$-graded Lie superalgebras and of their generalizations have been constantly investigated, see
\cite{{sch}, {sil},{SZZ}, {sil2}, {CSV},{sisi}, {cart},{AIS},{brdu2},{ISV},{Me}}. In physics ${\mathbb Z}_2\times{\mathbb Z}_2$-graded Lie superalgebras have been studied
in the contexts of de Sitter spaces \cite{{luri},{vas},{tol1}}, quasispin \cite{JYW}, strings \cite{zhe}, extension of Poincar\'e algebras \cite{{WT1},{WT2}}, double field theories \cite{BHPR}, mixed tensors \cite{{CKRS},{brib}}.\par  
More recent results concern symmetry. It has been shown in  \cite{{akttll1},{akttll2}} that
${\mathbb Z}_2\times{\mathbb Z}_2$-graded Lie superalgebras appear as symmetries of the L\'evy-Leblond Partial Differential Equations \cite{LL} describing nonrelativistic spin-$\frac{1}{2}$ particles. A single-particle quantum Hamiltonian dependent on a prepotential function and possessing a ${\mathbb Z}_2\times{\mathbb Z}_2$-graded symmetry has been introduced in \cite{brdu}. A systematic Lagrangian construction of ${\mathbb Z}_2\times{\mathbb Z}_2$-graded invariant models of classical mechanics has been presented in \cite{aktclass}. The theories described in \cite{aktclass} possess four types of particles: ordinary bosons, exotic bosons and two types of fermions (two fermions of different type mutually commute instead of anticommuting).  The canonical quantization of some of these models has been performed in \cite{aktquant}. The resulting Hamiltonians possess a ${\mathbb Z}_2\times{\mathbb Z}_2$-graded invariance. Besides recovering the \cite{brdu} single-particle quantum Hamiltonians, in \cite{aktquant} multiparticle Hamiltonians which allow the presence of interacting terms have also been obtained.  The most recent work on this line of research concerns the construction of two-dimensional sigma-models \cite{brusigma}.\par
Parastatistics were introduced in\cite{gre} by replacing the ordinary canonical (anti)commutation relations with more general algebraic triple relations. Triple relations also appear in the systems of combined parabosons and parafermions investigated in \cite{grme}. It was pointed out in\cite{gapa} (for parabosons)
and \cite{pal} (for parabosons and parafermions) that triple relations can be realized as  graded Jacobi identities of certain Lie superalgebras.  In the light of these results the theory of Lie superalgebras and their representations finds application to parastatistics. The specific type of ${\mathbb Z}_2\times{\mathbb Z}_2$-graded parastatistics was first introduced in \cite{{yaji1},{yaji2}}. In a series of papers \cite{{kd1},{kaha},{kaha2},{kan}} the ${\mathbb Z}_2\times{\mathbb Z}_2$-graded parastatistics was investigated in a Hopf algebra framework (a viewpoint which is also adopted in this work). The most recent papers on ${\mathbb Z}_2\times{\mathbb Z}_2$-graded parastatistics are \cite{{tol2},{stvdj}}.\par
Just like ordinary boson\slash fermion statistics does not require a supersymmetric theory to be applied, similarly the ${\mathbb Z}_2\times{\mathbb Z}_2$-graded parastatistics only requires the particles to be consistently accommodated into a ${\mathbb Z}_2\times{\mathbb Z}_2$-graded setting. Nevertheless, the presence of a ${\mathbb Z}_2\times{\mathbb Z}_2$-graded symmetry (which necessarily implies a ${\mathbb Z}_2\times{\mathbb Z}_2$-graded setting) simplifies the analysis of the problem. \par
The quantum models in \cite{{brdu},{aktquant}} are
${\mathbb Z}_2\times{\mathbb Z}_2$-graded invariants (under both the
${\mathbb Z}_2\times{\mathbb Z}_2$-graded one-dimensional Poincar\'e  and the Beckers-Debergh \cite{bede} algebras). The simplest of these models is a $4\times 4$ matrix oscillator Hamiltonian. It is the case investigated here
since it allows being analyzed with the powerful Hopf algebra tools (the noninteracting multiparticle states being constructed from coproducts, see also the approaches in \cite{{cachto},{cckt},{kuztop}}), within the framework described in Chapter $9$ of \cite{maj}: the braid statistics is encoded in a braided tensor product. For the case at hand the braiding is simply given by signs induced by the ${\mathbb Z}_2\times{\mathbb Z}_2$-grading. \par
It should be stressed that, up to our knowledge,  this is the first paper in which
the consequences of ${\mathbb Z}_2\times{\mathbb Z}_2$-graded parastatistics are tested in a Hamiltonian  theory. The papers \cite{{yaji1},{yaji2},{kd1},{kaha},{kaha2},{kan},{tol2},{stvdj}} are theoretical in nature and skip this point. The construction of a ${\mathbb Z}_2\times{\mathbb Z}_2$-graded Hamiltonian as an open problem to be left for future investigations was mentioned in \cite{kan}. \par
Concerning the structure of the paper, before addressing the ${\mathbb Z}_2\times{\mathbb Z}_2$-graded $4\times 4$ matrix oscillator (which can also be regarded as a ${\cal N}=4$ model of supersymmetric quantum mechanics \cite{wit}),
we discuss at first for propaedeutic reasons the $2\times 2$ matrix oscillator with ${\cal N}=2$ extended supersymmetries. Following a modern reinterpretation \cite{cachto} of the Wigner's quantization \cite{wig} this model can be regarded either as a bosonic theory (solved by a spectrum-generating algebra of ordinary creation/annihilation operators) or, alternatively, as a supersymmetric theory whose spectrum is recovered from a lowest weight representation of a Lie superalgebra. This is the content of the algebra/superalgebra duality discussed in \cite{top1}. At this level the $2\times 2$ oscillator admits in the single-particle sector two interpretations (bosonic and supersymmetric); they become two
inequivalent variants of the theory in the multiparticle sector. Built on that, the $4\times 4$ matrix oscillator admits three interpretations (bosonic, supersymmetric and ${\mathbb Z}_2\times{\mathbb Z}_2$-graded) as a single particle model. These three interpretations generate three {\it inequivalent} models (bosonic, supersymmetric and ${\mathbb Z}_2\times{\mathbb Z}_2$-graded) for the {\it multiparticle} theory. The main results are presented in Section {\bf 6}.\par
 \par
The scheme of the paper is the following. Section {\bf 2} presents the two interpretations 
for the single-particle Hamiltonian associated with the ${\cal N}=2$ supersymmetric oscillator. In Section {\bf 3} the three interpretations  of the single-particle  ${\cal N}=4$ supersymmetric oscillator are introduced. The two inequivalent   multiparticle systems associated with the ${\cal N}=2$ oscillator are presented in Section {\bf 4}. The three different multiparticle systems (bosonic, supersymmetric and ${\mathbb Z}_2\times {\mathbb Z}_2$-graded) associated with the ${\cal N}=4$ oscillator are computed in Section {\bf 5}. The proof that the ${\mathbb Z}_2\times{\mathbb Z}_2$-graded and
supersymmetric versions of the multiparticle systems lead to inequivalent theories is given in Section {\bf 6}.  A summary of results with comments and directions of future investigations is presented in Conclusions. Two appendices are included in order to make the paper self-contained. In Appendix {\bf A} the basic features of ${\mathbb Z}_2\times{\mathbb Z}_2$-graded Lie superalgebras are briefly recalled, while in Appendix {\bf B} a summary of Hopf algebras and braided tensor products is furnished.
 
\section{Two graded variants of the ${\cal N}=2$ supersymmetric oscillator}

In this Section we consider the $2\times 2$ matrix Hamiltonian $H_{2}$ of the  one-particle ${\cal N}=2$ supersymmetric oscillator, given by

\bea\label{h2osc}
H_{2} &=& {\footnotesize{\frac{1}{2}\left(
\begin{array}{cc} -\partial_x^2+x^2-1&0\\0&-\partial_x^2+x^2+1
\end{array}
\right)}}.
\eea
It is expressed as 
\bea
H_{2} &=&B_2^\dagger B_2 + f^\dagger f,  \qquad{\textrm{with}}\quad  f^\dagger f= {\footnotesize{\left(
\begin{array}{cc} 0&0\\0&1
\end{array}
\right)}}
,
\eea
in terms of the operators $B_2,B_2^\dagger,f,f^\dagger$ given by
\bea\label{bbff}
&B_2=b\cdot {\mathbb I}_2, \quad B_2^\dagger = b^\dagger\cdot{\mathbb I}_2,\quad{\textrm{with}}\quad b = \frac{i}{\sqrt 2}(\partial_x+x),\quad b^\dagger= \frac{i}{\sqrt 2}(\partial_x-x),
&\nonumber\\
& f = 
{\footnotesize{ \left(
\begin{array}{cc} 0&1\\0&0
\end{array}
\right)}}
, \quad f^\dagger ={\footnotesize{
 \left(
\begin{array}{cc} 0&0\\1&0
\end{array}\right)}}.
&
\eea
The creation, $B_2^\dagger,f^\dagger$, and annihilation, $B_2,f$, operators satisfy the commutators
\bea
&[H_{2},B_2]=-B_2, \quad [H_{2},B_2^\dagger]=B_2^\dagger, \quad,[H_{2}, f]=-f, \quad [H_{2},f^\dagger]=f^\dagger.&
\eea
The normalized Fock vacuum state $|vac\rangle_{2}$ is introduced through the positions
\bea\label{fock2osc}
&B_2|vac\rangle_{2}=f|vac\rangle_{2}=0, \qquad (\lVert|vac\rangle_{2}\rVert =1), \quad {\textrm{so that}}&
\eea
\bea\label{vac2osc}
|vac\rangle_{2}&=& {\footnotesize{ \pi^{-\frac{1}{4}}\left(\begin{array}{c} \exp{(-\frac{1}{2}x^2)}\\0\end{array}\right).}}
\eea
The single particle Hilbert space ${\cal H}_{2}^{(1)}$ is spanned by the normalized energy eigenvectors
$|n;\delta\rangle_{2}$, introduced through
\bea\label{spanhilb2osc}
|n;\delta\rangle_{2} &=& \frac{1}{\sqrt{n!}}(B_2^\dagger)^n (f^\dagger)^\delta|vac\rangle_{2}, \qquad {\textrm{for}} \quad n\in {\mathbb N}_0,\quad \delta=0,1.
\eea
They are such that
\bea
H_{2} |n;\delta\rangle_{2} &=& (n+\delta) |n;\delta\rangle_{2}, \qquad (
|vac\rangle_{2}\equiv |0;0\rangle_{2}).\eea
The energy spectrum 
\bea 
~~E_{n,\delta}&=& n+\delta,\qquad\quad \quad  E_{n+\delta}\in {\mathbb N}_0,
\eea is doubly degenerate for positive integers since
\bea
\qquad\quad  E_{n-1,1}&=&E_{n,0}=n\qquad \quad {\textrm{for}}\quad n=1,2,\ldots .
\eea
This degeneracy is denoted as ``$d_2(E)$". It follows that $d_2(0)=1$, $d_2(n)=2$ for $n\in{\mathbb N}$.\par
The two hermitian operators $Q_1, Q_2$ ($Q_i^\dagger=Q_i$ for $i=1,2$), given by
\bea
Q_1 = B_2f^\dagger+fB_2^\dagger, && Q_2 =-iB_2f^\dagger +ifB_2^\dagger,
\eea
satisfy the ${\cal N}=2$ one-dimensional super-Poincar\'e algebra
\bea
\relax \{Q_i,Q_j\} = 2\delta_{ij}H_{2}, &&  [H_{2}, Q_i]=0 \quad({\textrm{for}}\quad i,j=1,2).
\eea
Due to the presence of $Q_1,Q_2$, the Hamiltonian $H_{2}$ is ${\cal N}=2$ supersymmetric. This property explains \cite{wit} the double degeneracy of the energy spectrum  for positive integer  eigenvalues.\par
The single-particle quantum theory defined by the $H_{2}$ Hamiltonian admits two different, but physically equivalent, interpretations: a bosonic and a supersymmetric interpretation. From now on, following the conventions introduced in Appendix {\bf A}, a given ${\mathbb Z}_2^{[p]}$-graded Lie (super)algebra (with $p=0,1,2$)  will be denoted with the symbol ``${\mathfrak g}^{[p]}$" to stress its graded properties.\par
~\par
 Let us now  briefly discuss the two interpretations.\par
~\\
{\it Interpretation} $1$ ({\it bosonic}): in this  version of the theory all orthonormal states $|n;\delta\rangle_{2}$ spanning the Hilbert space ${\cal H}_{2}^{(1)}$ are assumed to be bosonic. The creation/annihilation operators 
$B_2,B_2^\dagger,f,f^\dagger$ are generators of an ordinary Lie algebra defined by commutators. The spectrum-generating Lie algebra ${\mathfrak{L}_{2}^{[0]}}$ of the model is spanned by the $6$ generators ${\mathbb I}_2,B_2,B_2^\dagger, f,f^\dagger,h$. We have
\bea\label{bosonicspectrumgen}
\{{\mathbb I}_2,B_2,B_2^\dagger, f,f^\dagger,h\}\in  {\mathfrak{L}_{2}^{[0]}}, &{\textrm{where}}& {\mathfrak{L}}_{2}^{[0]}=\mathfrak{h}(1)\oplus {\mathfrak{su}}_2
\eea
(here and throughout the paper, the symbol ``${\mathbb I}_n$" denotes the $n\times n$ identity matrix).\par
The nonvanishing commutators in  ${\mathfrak{L}_{2}^{[0]}}$ are
\bea
[B_2,B_2^\dagger] ={\mathbb I}_2,&\qquad& [f,f^\dagger]=h, \quad [h,f]=2f, \quad [h,f^\dagger]=-2f^\dagger.
\eea
The generators $B_2,B_2^\dagger, {\mathbb I}_2$ belong to the Heisenberg subalgebra ${\mathfrak{h}}(1)$, while
$h$ is the Cartan element and $f,f^\dagger$ the positive/negative roots of ${\mathfrak{su}}_2$. The operators 
$h,f,f^\dagger$ close a spin-$\frac{1}{2}$ representation of ${\mathfrak{su}}_2$
with $f,f^\dagger$ given in  (\ref{bbff}), while $h$ is
\bea
h &=&{\footnotesize{
 \left(
\begin{array}{cc} 1&0\\0&-1
\end{array}\right)}}.
\eea
\par
~\\
{\it Interpretation} $2$ ({\it supersymmetric}): this is the ``original" interpretation of $H_{2}$ as a supersymmetric Hamiltonian. The energy eigenstates $|n;\delta\rangle_{2}$ are assumed to be bosonic for $\delta=0$ and fermionic for $\delta=1$. The creation/annihilation operators $B_2,B_2^\dagger,f,f^\dagger$ belong
to a ${\mathbb Z}_2$-graded Lie superalgebra, with $B_2,B_2^\dagger$ in the even sector and $f,f^\dagger$ in the odd sector. The spectrum-generating Lie superalgebra ${\mathfrak{L}}_{2}^{[1]}$ of the model  is spanned by the $5$ generators ${\mathbb I}_2,B_2,B_2^\dagger,f,f^\dagger$:
\bea
\{{\mathbb I}_2,B_2,B_2^\dagger, f,f^\dagger\}\in  {\mathfrak{L}_{2}^{[1]}}, \quad&\quad&{\textrm{where}}\quad  {\mathfrak{L}}_{2}^{[1]}=\mathfrak{h}(1|1),\nonumber\\
\{{\mathbb I}_2,B_2,B_2^\dagger \}\in \mathfrak{h}(1|1)_{[0]}\equiv \mathfrak{h}(1),&\quad &\{f,f^\dagger\}\in  \mathfrak{h}(1|1)_{[1]}.
\eea
The nonvanishing (anti)commutators defining $\mathfrak{h}(1|1)$ are
\bea
[B_2,B_2^\dagger] ={\mathbb I}_2,&\qquad& \{f,f^\dagger\}={\mathbb I}_2.
\eea
This is the Heisenberg superalgebra of one bosonic and one fermionic oscillator.\par
~\par
The following comments are relevant:
\\~\\
{\it 1$^{st}$ comment}: for the single-particle Hamiltonian the two interpretations {\it i}) and {\it ii}) are physically equivalent. They are both admissible interpretations of the same physical model.\par
~\\
{\it 2$^{nd}$ comment}: for the multiparticle Hamiltonians constructed from $H_{2}$ the situation changes.
The constructions {\it i}) and {\it ii}) are no longer just interpretations, but two different variants producing inequivalent physical models.  To understand how this is possible one should take into account  that indistinguishable particles must be properly (anti)-symmetrized in accordance with the boson/fermion statistics.\par
~\\
{\it 3$^{rd}$ comment}: as a technical remark, the inequivalence of the multiparticle theories can also be understood as follows. The condition $(f^\dagger)^2=0 $ is encoded  in the Lie superalgebra $\mathfrak{h}(1|1)$ itself as the anticommutator $\{f^\dagger,f^\dagger\}=0$, while in the bosonic version it is the output of the spin-$\frac{1}{2}$ representation of $\mathfrak{su}(2)$. Multiparticle-theories induce higher spin representations of ${\mathfrak{su}}(2)$.\par
~\\
{\it 4$^{th}$ comment}: in the supersymmetric interpretation the operator
\bea\label{nfop}
N_F&=&{\footnotesize{
 \left(
\begin{array}{cc} 0&0\\0&1
\end{array}\right)}}
\eea
is the ``Fermion Parity Operator", defining bosons (fermions) as its $0$ (respectively, $1$) eigenstates. It is related by a constant diagonal shift to the ``spin operator" $S$ of the bosonic interpretation, defined as
\bea\label{spinop}
&S= -\frac{1}{2}h ={\footnotesize{
 \left(
\begin{array}{cc} -\frac{1}{2}&0\\0&\frac{1}{2}
\end{array}\right)}}.&
\eea 

\section{Three graded variants of the ${\cal N}=4$ supersymmetric oscillator}

The $4\times 4$ matrix Hamiltonian $H_{4}$ of the one-particle ${\cal N}=4$ supersymmetric oscillator is 
\bea\label{ham4osc}
H_{4}&=& {\footnotesize{\frac{1}{2}\left(
\begin{array}{cccc} -\partial_x^2+x^2-1&0&0&0\\0&-\partial_x^2+x^2-1&0&0\\0&0&-\partial_x^2+x^2+1&0\\0&0&0&-\partial_x^2+x^2+1
\end{array}
\right)}}.
\eea
It corresponds to a ``doubling" of the ${\cal N}=2$ oscillator since
\bea\label{double}
H_{4} &\equiv& H_{2}\oplus H_{2}.
\eea
It follows, in particular, that its energy spectrum coincides with the $H_{2}$ energy spectrum, the energy eigenvalues being $E_n=n$ for $n\in{\mathbb N}_0$, but the degeneracy $d_4(E_n)$ of each energy level $E_n$ is twice with respect to $d_2(E_n)$:
\bea
H_{2}&:&\qquad   d_2(0) =1, \qquad d_2(n) =2 \quad {\textrm{for}}\quad n\in {\mathbb N},\nonumber\\
H_{4}&:&\qquad   d_4(0) =2, \qquad d_4(n) =4 \quad {\textrm{for}}\quad n\in {\mathbb N}.
\eea
The ${\cal N}=4$ supersymmetry is guaranteed by the existence of $4$ hermitian operators ${\overline Q}_I$
($I=1,2,3,4$ and $ {\overline Q}_I^\dagger={\overline Q}_I$) satisfying the one-dimensional ${\cal N}=4$ super-Poincar\'e algebra
\bea
\relax \{{\overline Q}_I,{\overline Q}_J\} = 2\delta_{IJ}H_{4}, &&  [H_{4}, {\overline Q}_I]=0 \quad({\textrm{for}}\quad I,J=1,2,3,4).
\eea
The supersymmetry operators ${\overline Q}_I$ can be expressed as
\bea
{\overline Q}_1={\footnotesize{\frac{i}{\sqrt{2}}\left(
\begin{array}{cccc} 0&0&-\partial_x+x&0\\0&0&0&\partial_x-x\\-\partial_x-x&0&0&0\\0&\partial_x+x&0&0
\end{array}
\right)}},&&{\overline Q}_2={\footnotesize{\frac{1}{\sqrt{2}}\left(
\begin{array}{cccc} 0&0&0&\partial_x-x\\0&0&-\partial_x+x&0\\0&\partial_x+x&0&0\\-\partial_x-x&0&0&0
\end{array}
\right)}},\nonumber\\
{\overline Q}_3={\footnotesize{\frac{i}{\sqrt{2}}\left(
\begin{array}{cccc} 0&0&0&-\partial_x+x\\0&0&-\partial_x+x&0\\0&-\partial_x-x&0&0\\-\partial_x-x&0&0&0
\end{array}
\right)}},&&{\overline Q}_4={\footnotesize{\frac{1}{\sqrt{2}}\left(
\begin{array}{cccc} 0&0&\partial_x-x&0\\0&0&0&\partial_x-x\\-\partial_x-x&0&0&0\\0&-\partial_x-x&0&0
\end{array}
\right)}}.\nonumber\\
&&
\eea
The Hamiltonian $H_{4}$ is also invariant (see \cite{{{brdu},{aktquant}}}) under the ${\mathbb Z}_2\times{\mathbb Z}_2$-graded one-dimensional Poincar\'e superalgebra ${\mathfrak p}^{[2]}$ defined by the operators $H_{4}, {\overline Q}_1, {\overline Q}_2, Z$
accommodated in the
 graded sectors
\bea
&H_{4}\in {\mathfrak p}^{[2]}_{[00]},\qquad \quad{\overline Q}_1\in {\mathfrak p}^{[2]}_{[10]},\qquad \quad
{\overline Q}_2\in {\mathfrak p}^{[2]}_{[01]},\qquad \quad Z\in {\mathfrak p}^{[2]}_{[11]}. &
\eea
The (anti)commutators definining ${\mathfrak p}^{[2]}$ are
\bea
[{\overline Q}_1,{\overline Q}_2] = iZ, \quad [Z,H_{4}]=0,&\quad& [H_{4},{\overline Q}_1]=[H_{4},{\overline Q}_2]=0,\nonumber\\
\{{\overline Q}_1,{\overline Q}_1\}=\{{\overline Q}_2,{\overline Q}_2\}=H_{4}, &\quad & \{Z,{\overline Q}_1\}=\{Z,{\overline Q}_2\}=0.
\eea
The hermitian operator $Z=Z^\dagger$ is
\bea
Z&=&{\footnotesize{
\left(
\begin{array}{cccc} 0&-\partial_x^2+x^2-1&0&0\\-\partial_x^2+x^2-1&0&0&0\\0&0&0&-\partial_x^2+x^2+1\\0&0&-\partial_x^2+x^2+1&0
\end{array}
\right)}}.
\eea
We can introduce the following creation, $B_4^\dagger, a_0^\dagger, a_1^\dagger, a_2^\dagger$, and annihilation,
 $B_4, a_0, a_1, a_2$, operators defined as
\bea\label{b4b4dagger}
&B_4^\dagger =b\cdot {\mathbb I}_4,\qquad B_4=b^\dagger\cdot {\mathbb I}_4, \qquad{\textrm{where}}\quad b = \frac{i}{\sqrt 2}(\partial_x+x),\quad b^\dagger= \frac{i}{\sqrt 2}(\partial_x-x)&
\eea
and
\bea\label{aadagger}
&a_0={\footnotesize{
\left(
\begin{array}{cccc} 0&1&0&0\\0&0&0&0\\0&0&0&0\\0&0&0&0
\end{array}
\right)}},
\quad a_1={\footnotesize{
\left(
\begin{array}{cccc} 0&0&1&0\\0&0&0&0\\0&0&0&0\\0&0&0&0
\end{array}
\right)}},\quad a_2={\footnotesize{
\left(
\begin{array}{cccc} 0&0&0&1\\0&0&0&0\\0&0&0&0\\0&0&0&0
\end{array}
\right)}},&\nonumber\\
&a_0^\dagger={\footnotesize{
\left(
\begin{array}{cccc} 0&0&0&0\\1&0&0&0\\0&0&0&0\\0&0&0&0
\end{array}
\right)}},
\quad a_1^\dagger={\footnotesize{
\left(
\begin{array}{cccc} 0&0&0&0\\0&0&0&0\\1&0&0&0\\0&0&0&0
\end{array}
\right)}},\quad a_2^\dagger={\footnotesize{
\left(
\begin{array}{cccc} 0&0&0&0\\0&0&0&0\\0&0&0&0\\1&0&0&0
\end{array}
\right)}}.&
\eea
We have 
\bea
&[H_{4},B_4]=-B_4, \quad [H_{4},B_4^\dagger]=B_4^\dagger, \quad,[H_{4}, a_i]=-a_i, \quad [H_{4},
a_i^\dagger]=a_i^\dagger \quad{\textrm{for}}\quad i=1,2,&\nonumber\\
\eea
while, due to the (\ref{double}) double degeneracy of the states,
\bea
[H_{4},a_0]= [H_{4},a_0^\dagger]&=&0.
\eea
The normalized Fock vacuum $|vac\rangle_{4}$ is annihilated by $B_4,a_0,a_1,a_2$:
\bea
&B_4|vac\rangle_{4}=a_0|vac\rangle_{4}=a_1|vac\rangle_{4}=a_2|vac\rangle_{4}=0,&
\eea
so that
\bea
|vac\rangle_{4}&=& {\footnotesize{ \pi^{-\frac{1}{4}} e^{(-\frac{1}{2}x^2)}\left(\begin{array}{c}1\\0\\0\\0\end{array}\right).}}
\eea
The single particle Hilbert space ${\cal H}_{4}^{(1)}$ is spanned by the normalized energy eigenvectors
$|n;\delta_0\delta_1\delta_2\rangle_{4}$, introduced through
\bea\label{n4basis}
|n;\delta_0\delta_1\delta_2\rangle_{4} &=& \frac{1}{\sqrt{n!}}(B_4^\dagger)^n (a_0^\dagger)^{\delta_0}(a_1^\dagger)^{\delta_1}(a_2^\dagger)^{\delta_2}|vac\rangle_{4}, \quad {\textrm{for}} \quad n\in {\mathbb N}_0,\quad \delta_0,\delta_1,\delta_2=0,1,\nonumber\\
&&{\textrm{with}}\quad 0\leq\delta_0+\delta_1+\delta_2\leq 1.
\eea
The constraint on $\delta_0+\delta_1+\delta_2$  is due to the relations, for the creation operators $a_0^\dagger, a_1^\dagger, a_2^\dagger$, 
\bea\label{rsnull}
\qquad\qquad  a_r^\dagger a_s^\dagger &=&0 \qquad\quad\forall ~r,s =0,1,2.
\eea
The energy eigenvalues $E_{n;\delta_0\delta_1\delta_2}$ are
\bea
H_{4} |n;\delta_0\delta_1\delta_2\rangle_{4} &=& E_{n;\delta_0\delta_1\delta_2} |n;\delta_0\delta_1\delta_2\rangle_{4},\nonumber\\ E_{n;\delta_0\delta_1\delta_2}&=&n+\delta_1+\delta_2.
\eea
The Fock state $|vac\rangle_{4}\equiv |0;000\rangle_{4}$ is one of the two degenerate vacua of the theory, the other vacuum state being $|0;100\rangle_{4}$.\par 

It is now clear that three interpretations for the Hamiltonian $H_{4}$ can be given.\par
~\\
{\it Interpretation} $1$ ({\it bosonic}): in this  version of the theory all states are bosonic.\par~\\
{\it Interpretation} $2$ ({\it supersymmetric}): the states satisfying  $\delta_1+\delta_2=0~({\textrm{mod}}~2)$ are bosons; the states satisfying $\delta_1+\delta_2=1$ are fermions.  The creation operators $B_4^\dagger,a_0^\dagger$ are even, while $a_1^\dagger, a_2^\dagger $ are odd.\par
~\\
{\it Interpretation} $3$ ({\it ${\mathbb Z}_2\times{\mathbb Z}_2$-graded)}: the Hilbert space ${\cal H}_{4}^{(1)}$ is ${\mathbb Z}_2\times{\mathbb Z}_2$-graded:
\bea
{\cal H}_{4}^{(1)}&=&{{\cal H}_{4}^{(1)}}_{[00]}\oplus{{\cal H}_{4}^{(1)}}_{[11]}\oplus{{\cal H}_{4}^{(1)}}_{[10]}\oplus{{\cal H}_{4}^{(1)}}_{[01]}.
\eea 
The states $|n;~\delta_0~\delta_1~\delta_2\rangle_{4}$, for any $n\in{\mathbb N}_0$, are accommodated into its graded sectors as
\bea
\{|n;000\rangle_{4}\}\in {{\cal H}_{4}^{(1)}}_{[00]},&&
\{|n;100\rangle_{4}\}\in{{\cal H}_{4}^{(1)}}_{[11]}, \nonumber\\
\{|n;010\rangle_{4}\}\in {{\cal H}_{4}^{(1)}}_{[10]}, &&
\{|n;001\rangle_{4}\}\in {{\cal H}_{4}^{(1)}}_{[01]}.
\eea
The ${\mathbb Z}_2\times{\mathbb Z}_2$-grading of the creation operators is
\bea
& B_4^\dagger \in [00],\qquad a_0^\dagger\in [11], \qquad  a_1^\dagger\in [10], \qquad a_2^\dagger\in [01].&
\eea
\par
~\\
{\it 1$^{st}$ comment}: the three interpretations given above produce three different variants with inequivalent physics for the respective multiparticle sectors of the quantum model.\par
~\\
{\it 2$^{nd}$ comment}: due to the relations (\ref{rsnull}) the creation operators $a_0^\dagger,a_1^\dagger,a_2^\dagger$ satisfy, in each of the three above respective cases, a graded abelian (super)algebra. The (anti)commutators are 
\bea [a_r^\dagger,a_s^\dagger\}&=&0.
\eea
The brackets  $[\cdot,\cdot\}$, see (\ref{mixedbracket}),  are defined in accordance with the respective grading.

\section{Multiparticle sectors of the ${\cal N}=2$ supersymmetric oscillator}

For propaedeutic reasons, before addressing the three multiparticle variants of the ${\cal N}=4$ supersymmetric oscillator, it is here illustrated the construction of multiparticle states, via coproduct and braided tensors, of the ${\cal N}=2$ supersymmetric oscillator in its two variants. The relevant formulas presented in Section {\bf 2} and Appendix {\bf B} are recalled. To illustrate the nuances of the construction that is later applied to the ${\mathbb Z}_2\times{\mathbb Z}_2$-graded oscillator, some heavy notation is carried out. At the end the results are summarized in a simplified notation.   \par
In the single-particle case the state of the system is uniquely determined by the two compatible observables  
$H_{2}$ ($H_{2}=B_2^\dagger B_2\cdot{\mathbb I}_2+N_F$) and $N_F$ introduced in (\ref{h2osc}) and (\ref{nfop}), respectively.  The unique bosonic vacuum
state $|vac\rangle_{2}$ introduced in (\ref{vac2osc}) is defined as satisfying the Fock's conditions (\ref{fock2osc}). The single-particle Hilbert space ${\cal H}_{2}^{(1)}$ is spanned, see (\ref{spanhilb2osc}), by the creation operators ${B_2}^\dagger, f^\dagger$ acting on the vacuum.\par
The $M>1$ multiparticle Hilbert space ${\cal H}_{2}^{(M)}$ is a subset of tensor products of $M$ single-particle Hilbert spaces:
\bea
\qquad {\cal H}_{2}^{(M)}\subset {\cal H}_{2}^{(1)}\otimes \ldots \otimes {\cal H}_{2}^{(1)}, && ({\textrm{tensor ~product~of~ $M$~ spaces}}).
\eea
The coproduct $\Delta:{\cal U}\rightarrow{\cal U}\otimes{\cal U}$ of a Universal Enveloping Algebra ${\cal U}(\mathfrak{g}^{[p]})$ of a graded Lie algebra $\mathfrak{g}^{[p]}$, see (\ref{costructures}),  satisfies  $\Delta({\mathbf 1})={\mathbf 1}\otimes{\mathbf 1}$, $\Delta(g)={\mathbf 1}\otimes g+g\otimes{\mathbf 1}$ for
$g\in \mathfrak{g}^{[p]}$ and $\Delta(U_AU_B)=\Delta(U_A)\Delta(U_B)$ for $U_{A,B}\in {\cal U}$  (formulas (\ref{deltaid}, \ref{deltag}, \ref{deltauu}), respectively). \par
The coassociativity property (\ref{coassociativity}) allows to recursively determine $\Delta^{(M+1)}$ as
\bea
\Delta^{(M+1)}&=& ({\mathbf 1}\otimes \Delta)\Delta^{(M)}=(\Delta\otimes {\mathbf 1})\Delta^{(M)} \qquad ({\textrm{with}}~ \Delta^{(1)}\equiv \Delta),
\eea
where $\Delta^{(M)}$ maps ${\cal U}$ in the tensor product of $M+1$ spaces:
\bea
\Delta^{(M)} &:& {\cal U} \rightarrow {\cal U}^{\otimes M+1}.
\eea
The braiding of the tensor spaces, defined in (\ref{epsilonbraiding}), for the cases under consideration here is at most realized by  a $-1$ sign.\par
The unique bosonic vacuum $|vac\rangle_{2}^{(M)}$ of the $M$-particle ${\cal N}=2$ oscillator is determined
by the Fock conditions, for the annihilation operators $B_2,f$ introduced in (\ref{bbff}),
\bea
\Delta^{(M-1)}(B_2) |vac\rangle_{2}^{(M)}= \Delta^{(M-1)}(f) |vac\rangle_{2}^{(M)}&=&0.
\eea
The normalized  $M$-particle vacuum $|vac\rangle_{2}^{(M)}$ is expressed
as
\bea
|vac\rangle_{2}^{(M)}&=&|vac\rangle_{2}\otimes \ldots \otimes |vac\rangle_{2}\in {\cal H}_{2}^{(M)}.
\eea
Once determined the $M$-particle vacuum, the $M$-particle excited states and the observables are recovered
from, respectively, the coproducts of the creation operators $B_2^\dagger, f^\dagger $ in (\ref{bbff}) and of 
$H_{2}, N_F$:
\bea
{\textrm{excited~states}}:\Delta^{(M-1)} ((B_2^\dagger)^n(f^\dagger)^r); &&  {\textrm{observables}}:   \Delta^{(M-1)} (H_{2}),\quad
\Delta^{(M-1)} (N_F),
\eea
where $n\in {\mathbb N}_0$ and $r$ is restricted as shown below. \par
We can therefore focus on the Universal Enveloping Algebras ${\cal U}$ induced by the operators $B_2^\dagger, f^\dagger, H_{2},N_f$.\par
 In the bosonic variant we have ${\cal U}({\mathfrak{t}}^{[0]})$, where ${\mathfrak{t}}^{[0]}$ is the Lie algebra defined by the
commutators
\bea\label{l0commutators}
&[H_{2},N_F]=[N_F,B_2^\dagger]=[B_2^\dagger, f^\dagger]=0, &\nonumber\\&[H_{2}, B_2^\dagger]=B_2^\dagger,\quad [H_{2}, f^\dagger]=f^\dagger,\quad  [N_F,f^\dagger]=f^\dagger.&
\eea

In the supersymmetric variant we have ${\cal U}({\mathfrak{t}}^{[1]})$, where ${\mathfrak{t}}^{[1]}$ is the Lie superalgebra recovered by assuming $H_{2},N_F,B_2^\dagger$ to be even and $f^\dagger$ to be odd. It is defined by
the same set of (\ref{l0commutators}) commutators with the addition of a single  (anti)commutator given by
\bea\label{extraanticomm}
\{f^\dagger,f^\dagger\}&=&0.
\eea
The  Lie superalgebra homomorphism realized by the coproduct implies, in the superalgebra case, that the extra
relation (\ref{extraanticomm}) produces, for any integer $M\geq 2$, the relations
\bea
\{\Delta^{(M-1)}(f^\dagger),\Delta^{(M-1)}(f^\dagger)\}=0 &\Rightarrow&\Delta^{(M-1)}((f^\dagger)^2)=0.
\eea
As explained in formula (\ref{fermionicfdagger}), the above equations are consequences of the $-1$ sign entering the braided tensor for the fermionic generator $f^\dagger$.\par
In the bosonic case we have $\Delta^{(M-1)}((f^\dagger)^2)\neq0$ for any integer $M\geq 2$ (see formula (\ref{bosonicfdagger}) for $M=2$). By taking into account that $(f^\dagger)^2=0$, the maximal value $r$ such that $\Delta^{(M-1)}((f^\dagger)^r)\neq0$ while $ \Delta^{(M-1)}((f^\dagger)^{r+1})=0$, is reached for $r=M$. For that value
\bea
\Delta^{(M-1)}((f^\dagger)^{M})&\propto& f^\dagger\otimes f^\dagger\otimes \ldots \otimes f^\dagger.
\eea
In the bosonic case the nonvanishing coproducts
\bea
\Delta^{(M-1)}((f^\dagger)^{r}), && r=0,1,\ldots, M,
\eea
are accommodated into a spin-$\frac{M}{2}$ representation of the $\mathfrak{su_2}$ subalgebra of the spectrum-generating algebra ${\mathfrak{L}_{2}^{[0]}}$ introduced in (\ref{bosonicspectrumgen}).\par
We can discriminate the bosonic versus the supersymmetric construction by introducing the $p=0,1$
parameter which denotes the ${\mathbb Z}_2^p$-grading of the ${\mathfrak t}^{[p]}$ (super)algebras introduced above.
By using this convention, the normalized $M$-particle states are expressed as
\bea
|n; r\rangle_p^{(M)} &\propto& {\widehat{ \Delta_p^{(M-1)} }((B_2^\dagger)^n(f^\dagger)^r)}|vac\rangle_{2}^{(M)}
\eea
(the hat indicates the evaluation of the coproducts in the given representation).\par
They are eigenstates
of the $M$-particle energy and $N_F$ operators, respectively denoted as 
${\widehat{\Delta_p^{(M-1)} }}(H_{2})$ and ${\widehat{\Delta_p^{(M-1)}}} (N_{F})$. Their eigenvalues are
\bea
{\widehat{\Delta_p^{(M-1)} }}(H_{2})|n; r\rangle_{p}^{(M)}&=&(n+r)|n; r\rangle_{p}^{(M)},\nonumber\\
{\widehat{\Delta_p^{(M-1)}}} (N_{F})|n; r\rangle_{p}^{(M)}&=&r|n; r\rangle_{p}^{(M)}.
\eea
The bosonic ($p=0$) and the supersymmetric ($p=1$) Hilbert spaces ${\cal H}_{p}^{(M)}$ are spanned by the eigenvectors
\bea\label{n2span}
|n; r\rangle_0^{(M)}&\in& {\cal H}_{0}^{(M)}\qquad {\textrm{with}}\quad n\in {\mathbb N}_0, \quad r=0, 1,\ldots, M,\nonumber\\
|n; r\rangle_1^{(M)}&\in&{\cal H}_{1}^{(M)}\qquad {\textrm{with}}\quad n\in {\mathbb N}_0, \quad r=0, 1.\nonumber\\
\eea
The supersymmetric Hilbert space is a subset of the bosonic Hilbert space,
\bea
&{\cal H}_{1}^{(M)}\subset {\cal H}_{0}^{(M)},&\nonumber\\
&{\textrm{since}}\quad |n; r\rangle_1^{(M)}=|n; r\rangle_0^{(M)}\quad{\textrm{in the common range}}\quad n\in {\mathbb N}_0,\quad r=0,1.&
\eea
For the $M=1$ single-particle case we have, in particular,
\bea
&|n; r\rangle_0^{(1)}=|n; r\rangle_1^{(1)}\equiv |n; r\rangle,\quad {\textrm{with}}\quad 
{\cal H}_{1}^{(1)}= {\cal H}_{0}^{(1)}.&
\eea
By taking into account that the creation/annihilation operators  $b = \frac{i}{\sqrt 2}(\partial_x+x), ~ b^\dagger= \frac{i}{\sqrt 2}(\partial_x-x)$ introduced in (\ref{bbff}) can be expressed, in each $j$-th space of the tensor product, by a different coordinate $x_j$ (the position of the $j$-th particle, with $x_jx_k=x_kx_j$ for any $j,k$), the derived multiparticle Hamiltonians $H^{(M)}$ and the diagonal operators $N_F^{(M)}= \Delta^{(M-1)}(N_F)$,
up to $M\leq 3$, can be written as
\bea
H^{(1)}&=& \frac{1}{2}(-\partial_{x_1}^2+x_1^2-1)\cdot {\mathbb I}_2 + N_F^{(1)},\qquad ~ \qquad\qquad\qquad\quad\quad N_F^{(1)}=diag(0,1),\nonumber\\
H^{(2)}&=& \frac{1}{2}(-\partial_{x_1}^2-\partial_{x_2}^2+x_1^2+x_2^2-2)\cdot {\mathbb I}_4 + N_F^{(2)},\qquad\qquad
~~ \quad N_F^{(2)}=diag(0,1,1,2),\nonumber\\
H^{(3)}&=& \frac{1}{2}(-\partial_{x_1}^2-\partial_{x_2}^2-\partial_{x_3}^2+x_1^2+x_2^2+x_3^2-3)\cdot {\mathbb I}_8 + N_F^{(3)},\quad N_F^{(3)}=diag(0,1,2,3,1,2,3,4).\nonumber\\
&&
\eea
In terms of the normalized  two-component single-particle states
\bea
|n;r\rangle &\equiv & \psi_{n;r} (x_1), \qquad r=0,1,
\eea 
the normalized $M=2$ multiparticle states $|n;r\rangle^{(2)}$ read as follows:
\bea
|n;0\rangle^{(2)}&=&N_{n;0}{\footnotesize{\sum_{k=0}^n \left(\begin{array}{c}n\\k\end{array}\right)}}|n-k;0\rangle \otimes |k;0\rangle\equiv\nonumber\\
&&N_{n;0}{\footnotesize{\sum_{k=0}^n \left(\begin{array}{c}n\\k\end{array}\right)}}\psi_{n-k;0}(x_1)\cdot \psi_{k;0}(x_2),\nonumber\\
|n;1\rangle^{(2)}&=& N_{n;1}{\footnotesize{\sum_{k=0}^n \left(\begin{array}{c}n\\k\end{array}\right)}}
(|n-k;1\rangle \otimes |k;0\rangle+|n-k;0\rangle \otimes |k;1\rangle)\equiv
\nonumber\\
&&N_{n;1}{\footnotesize{\sum_{k=0}^n \left(\begin{array}{c}n\\k\end{array}\right)}}(\psi_{n-k;1}(x_1)\cdot \psi_{k;0}(x_2)+\psi_{n-k;0}(x_1)\cdot \psi_{k;1}(x_2)),\nonumber\\
~^{(\ast)}~|n;2\rangle^{(2)}&=&  N_{n;2} {\footnotesize{\sum_{k=0}^n \left(\begin{array}{c}n\\k\end{array}\right)}}|n-k;1\rangle \otimes |k;1\rangle\equiv\nonumber\\
&&N_{n;2}{\footnotesize{\sum_{k=0}^n \left(\begin{array}{c}n\\k\end{array}\right)}}\psi_{n-k;1}(x_1)\cdot \psi_{k;1}(x_2).
\eea
The asterisk in front of the $|n;2\rangle^{(2)}$ states indicates that they only enter the bosonic theory. They are excluded in the supersymmetric theory due to the Pauli exclusion principle for fermions. \par
The normalization coefficients $N_{n;r}$ for $r=0,1,2$ are recovered from the central binomial coefficients $B_n=\frac{2n!}{(n!)^2}$ which give rise to the $1,2,6,20,70, 252, \ldots$ sequence (sequence $A000984$ in the OEIS, On-line
Encyclopedia of Integer Sequences, database). We have
\bea
N_{n;0}=N_{n;2} = \frac{1}{\sqrt{B_n}},&&
N_{n;1} = \frac{1}{\sqrt{2B_n}}.
\eea
The $M=3$ three-particle states $|n;r\rangle^{(3)}$ for $n>0$ can be easily recovered from the $n=0$, $|0;r\rangle^{(3)}$ normalized states given by
\bea
|0;0\rangle^{(3)}&=& |0;0\rangle\otimes|0;0\rangle\otimes |0;0\rangle,\nonumber\\
|0;1\rangle^{(3)}&=&\frac{1}{\sqrt{3}}(|0;1\rangle\otimes|0;0\rangle\otimes |0;0\rangle+|0;0\rangle\otimes|0;1\rangle\otimes |0;0\rangle+|0;0\rangle\otimes|0;0\rangle\otimes |0;1\rangle),\nonumber\\
~^{(\ast)}~|0;2\rangle^{(3)}&=&\frac{1}{\sqrt{3}}(|0;0\rangle\otimes|0;1\rangle\otimes |0;1\rangle+|0;1\rangle\otimes|0;0\rangle\otimes |0;1\rangle+|0;1\rangle\otimes|0;1\rangle\otimes |0;0\rangle), \nonumber\\
~^{(\ast)}~|0;3\rangle^{(3)}&=&|0;1\rangle\otimes|0;1\rangle\otimes |0;1\rangle.
\eea
The states $|0;2\rangle^{(3)},~|0;3\rangle^{(3)}$ only exist in the bosonic theory.
\subsection{Summary of results}
The $M$-particle energy eigenvalues $E_n$ coincide in both bosonic and supersymmetric variants,
\bea
E_n&=& n \quad {\textrm{with}}\quad n\in {\mathbb N}_0,
\eea
but their degeneracies differ. \par
For the supersymmetric theory the degeneracy $d_{susy}(E_n)$ is
\bea
\forall M=1,2,3,\ldots:&& d_{susy}(E_0) =1, \qquad d_{susy}(E_n)=2\quad {\textrm{for}}\quad n>0.
\eea
For the bosonic theory the degeneracy $d_{bos}(E_n)$ is
\bea
&& d_{bos}(E_n) =n+1 \quad {\textrm{for}}\quad n<M, \qquad d_{bos}(E_n)=M+1\quad {\textrm{for}}\quad n\geq M.
\eea
The above counting is understood by taking into account that for the bosonic states $|n;r\rangle^{(M)}$, which have energy eigenvalues $n+r$, $r$ can be at most $r=M$. Therefore, as an example, for the $M=2$ two-particle case, the $E=1$ energy eigenstates are 
$|1;0\rangle^{(2)}$ and $|0;1\rangle^{(2)}$. Starting from $E=2$ we get degenerate eigenstates which include $r=2$. The three degenerate states are
$|2;0\rangle^{(2)}$, $|1;1\rangle^{(2)}$ and $|0;2\rangle^{(2)}$. This counting is repeated for $E>2$, so that at $E=3$ we get 
$|3;0\rangle^{(2)}$, $|2;1\rangle^{(2)}$ and $|1;2\rangle^{(2)}$ and so on.\par
The following table illustrates the degeneracies up to $E=5$ in the supersymmetric and the bosonic (up to $M=5$ particles) cases:
\bea&
\begin{array}{|c||c|c|c|c|c|c|}\hline  E:&$0$&$1$&$2$&$3$ &$4$&$5$\\ \hline \hline
d_{susy}~ (\forall M):&1&2&2&2&2&2\\ \hline \hline
d_{bos} ~ (M=1):&1&2&2&2&2&2\\  \hline 
d_{bos} ~ (M=2):&1&2&3&3&3&3\\ \hline 
d_{bos} ~ (M=3):&1&2&3&4&4&4\\  \hline 
d_{bos} ~ (M=4):&1&2&3&4&5&5\\ \hline
d_{bos} ~ (M=5):&1&2&3&4&5&6\\ \hline
\end{array}&
\eea
The degenerate bosonic energy eigenstates are accommodated into 
\bea
&{\textrm{the spin-$\frac{n}{2}$ representation of ${\mathfrak{su_2}}$ for $n<M$}},&\nonumber\\
&{\textrm{the spin-$\frac{M}{2}$ representation of ${\mathfrak{su_2}}$ for $n\geq M$}}.&
\eea
The  operator $S$ which measures the spin, introduced in (\ref{spinop}), is a shift of the $N_F$ operator:
$S= N_F-\frac{1}{2}\cdot{\mathbb I}_2$.\par~
\par
{\it Comment}: a physical application can be envisaged based on the following scenario. Let's suppose that we have established that the single-particle sector of a physical system under investigation possesses the $E=0,1,2,3,\ldots$ energy spectrum with respective degeneracies $d(E)=1,2,2,2,\ldots$. We still don't know whether the system under consideration is only composed by bosons or  (the supersymmetric case being realized) by bosons and fermions. To establish this the investigation of the multiparticle sector is required. The simplest answer is provided for two particles at the energy level $E=2$. In the bosonic case the extra state $|0;2\rangle^{(2)}$, not allowed in the supersymmetric case, is present. The complete set of two-particle observables $H_2^{(2)},N_F^{(2)}$ uniquely characterize 
$|0;2\rangle^{(2)}$ by their pair of $(2,2)$ respective eigenvalues. If this pair of eigenvalues is observed, then the supersymmetric case must be excluded.

\section{Multiparticle sectors of the ${\cal N}=4$ supersymmetric oscillator}

We start here the investigation of the physical consequences, in the multiparticle sector, of the three graded variants  of the
${\cal N}=4$ supersymmetric oscillator introduced in (\ref{ham4osc}). The methods that we are employing have been illustrated in Section {\bf 4} in application to the ${\cal N}=2$ supersymmetric oscillator. This gives the opportunity to focus on the relevant questions and results without unnecessary distracting technicalities.\par
Since the main question that we are addressing concerns the physical signature of the ${\mathbb Z}_2\times{\mathbb Z}_2$-grading with respect to the ${\mathbb Z}_2$-grading, the bosonic variant will just be sketched (it will be discussed, in any case, for completeness). It is already clear, by extending the results of Section {\bf 4}, that it produces a different physics with respect to both other (${\mathbb Z}_2$- and ${\mathbb Z}_2\times{\mathbb Z}_2$-graded) variants of the theory. 

\subsection{The single-particle case revisited}

For our purposes we introduce two different complete sets of $4$ compatible observables.\par
The Hamiltonian (\ref{ham4osc}) can be written as
\bea\label{h4nf}
H_4 &=& \frac{1}{2}(-\partial_x^2+x^2-1)\cdot{\mathbb I}_4+N_f,\qquad {\textrm{with}} \quad N_f=
 {\footnotesize{\left(
\begin{array}{cccc} 0&0&0&0\\0&0&0&0\\0&0&1&0\\0&0&0&1
\end{array}
\right)}}.
\eea
With an abuse of language we can refer to $N_f$ as the Fermion Parity Operator, even if this is only true in the supersymmetric and ${\mathbb Z}_2\times{\mathbb Z}_2$-graded interpretations of the model.\par
The other observables that we are considering are
\bea\label{nvnw}
&N_v=
 {\footnotesize{\left(
\begin{array}{cccc} 0&0&0&0\\0&1&0&0\\0&0&0&0\\0&0&0&1
\end{array}
\right)}}, \qquad N_w=
 {\footnotesize{\left(
\begin{array}{cccc} 0&0&0&0\\0&1&0&0\\0&0&1&0\\0&0&0&0
\end{array}
\right)}},\qquad N_T=N_v+N_w&
\eea
and 
\bea\label{xf}
X_f&=&
 {\footnotesize{\left(
\begin{array}{cccc} 0&0&0&0\\0&0&0&0\\0&0&0&1\\0&0&1&0
\end{array}
\right)}}.
\eea
The operator $X_f$ exchanges the fermions in the ${\mathbb Z}_2$- and ${\mathbb Z}_2\times{\mathbb Z}_2$-graded interpretations of the model.\par
With these positions we have\\
~\par
{\it Set $1$}: the operators $H_4,N_f,N_v,N_w$ furnish a complete set of $4$ compatible, mutually commuting, observables;\\
~\par
{\it Set $2$}: the operators $H_4,N_f,N_T,X_f$ furnish a complete set of $4$ compatible, mutually commuting, observables.\\
~\par
The grading of these operators is summarized as follows.\\
~\par
{\it Remark $1$}: the diagonal operators $H_4,N_f,N_v,N_w,N_T$ are $0$-graded when assuming the ${\mathbb Z}_2^{0}$- and ${\mathbb Z}_2^{1}$-graded versions of the theory; they are $00$-graded in the ${\mathbb Z}_2^{2}$-graded version.
\\
~\par
{\it Remark $2$}: the operator $X_f$ is $0$-graded (bosonic) when assuming the ${\mathbb Z}_2^{0}$- and ${\mathbb Z}_2^{1}$-graded versions of the theory; it is $11$-graded in the ${\mathbb Z}_2^{2}$-graded version.\par
~\par
The single-particle Hilbert space from now on will be simply denoted as ${\cal H}^{(1)}$; it is spanned by the
orthonormal vectors $|n;\delta_0\delta_1\delta_2\rangle_4$ introduced in (\ref{n4basis}), with $n\in N_0$ and 
$ \delta_0,\delta_1,\delta_2$ satisfying $ 0\leq\delta_0+\delta_1+\delta_2\leq 1$. For simplicity we drop the suffix $4$ so that we can write
\bea\label{singlen4vectors}
\{|n;000\rangle, ~ |n;100\rangle, ~|n;010\rangle, ~|n;001\rangle \}&\in&{\cal H}^{(1)}, \qquad {\textrm{for}} \quad n\in {\mathbb N}_0.
\eea
As a vector space the Hilbert space ${\cal H}^{(1)}$ is the same in all three ${\mathbb Z}_2^p$, $p=0,1,2$, graded versions of the theory. Once introduced the symbol  ${\cal H}_p^{(M)}$ to denote the $M$-particle Hilbert space in the ${\mathbb Z}_2^p$-graded variant of the theory, the above statement can be expressed as
\bea
{\cal H}_0^{(1)}={\cal H}_1^{(1)}={\cal H}_2^{(1)}&\equiv & {\cal H}^{(1)}.
\eea

The vectors $|n;000\rangle, ~ |n;100\rangle, ~|n;010\rangle, ~|n;001\rangle $ form an orthonormal basis of eigenvectors for the 
$H_4,N_f,N_v,N_w$ set of complete compatible observables.\par
The vectors $|n;000\rangle,~ |n;100\rangle,~|n;0+\rangle,  ~ |n;0-\rangle$, with 
$|n;0\pm\rangle $ defined as
\bea
|n;0\pm\rangle &=&\frac{1}{\sqrt{2}}(|n;010\rangle\pm |n;001\rangle),
\eea
form an orthonormal basis of eigenvectors for the 
$H_4,N_f,N_T,X_f$ set of complete compatible observables.\par
The corresponding eigenvalues can be read from the respective tables below.\par
For the first set of observables we have:
\bea&
\begin{array}{|c|c|c|c|c|c|c|}\hline  &H_4&N_f&N_v&N_w\\ \hline 
|n;000\rangle:&n&0&0&0\\ \hline 
|n;100\rangle:&n&0&1&1\\ \hline
|n;010\rangle:&n+1&1&0&1\\ \hline 
|n;001\rangle:&n+1&1&1&0\\ \hline 
\end{array}&
\eea
For the second set of observables we have:
\bea&
\begin{array}{|c|c|c|c|c|c|c|}\hline  &H_4&N_f&N_T&X_f\\ \hline 
|n;000\rangle:&n&0&0&0\\ \hline 
|n;100\rangle:&n&0&2&0\\ \hline
|n;0+\rangle:&n+1&1&1&1\\ \hline 
|n;0-\rangle:&n+1&1&1&-1~~\\ \hline 
\end{array}&
\eea

\subsection{The multiparticle Hilbert spaces}

We present now the construction of the ${\mathbb Z}_2^p$-graded $M$-particle Hilbert spaces ${\cal H}_p^{(M)}$
for  $p=0,1,2$; its vectors will be denoted as $|v\rangle_p^{(M)}\in {\cal H}_p^{(M)}$. We have
\bea\label{subset}
 {\cal H}_p^{(M)}&\subset &({{\cal H}^{(1)}})^{\otimes M}.
\eea
The single-particle annihilation and creation operators have been introduced in (\ref{b4b4dagger})  ($B_4,B_4^\dagger $) and (\ref{aadagger}) ($a_0,a_1,a_2,a_0^\dagger,a_1^\dagger,a_2^\dagger$). The ${\mathbb Z}_2^p$-graded coproducts will be denoted as $\Delta_p$. Just like the ${\cal N}=2$ counterpart, the $M$-particle Fock vacuum $|0;000\rangle_p^{(M)}$ is defined by requiring
\bea
\Delta_p^{(M-1)}(g)|0;000\rangle_p^{(M)}&=&0\qquad{\textrm{for}}\quad g=B_4,a_0,a_1,a_2.
\eea
The same Fock vacuum $|0;000\rangle^{(M)}\equiv|0;000\rangle_p^{M)}$ is found for $p=0,1,2$; we have 
\bea
|0;000\rangle^{(M)}&=& |0;000\rangle\otimes\ldots \otimes |0;000\rangle~\in {\cal H}_p^{(M)}.
\eea
The excited states are created via the coproducts $\Delta_p$ defined on the Universal Enveloping Algebra
${\cal U}({\mathfrak a}^{[p]})$ of the ${\mathbb Z}_2^p$-graded abelian (i.e., all (anti)commutators are vanishing) superalgebra ${\mathfrak a}^{[p]}$ induced by the creation operators $B_4^\dagger, a_0^\dagger,a_1^\dagger,a_2^\dagger$. The grading assignment, see (\ref{gradingdecomposition}), is
\bea
p=0 &:& B_4^\dagger, a_0^\dagger,a_1^\dagger,a_2^\dagger\in {\mathfrak a}_{[0]}^{[0]},\nonumber\\
p=1 &:& B_4^\dagger, a_0^\dagger\in {\mathfrak a}_{[0]}^{[1]},\quad a_1^\dagger,a_2^\dagger\in {\mathfrak a}_{[1]}^{[1]},\nonumber\\
p=2 &:& B_4^\dagger\in {\mathfrak a}_{[00]}^{[2]},\quad  a_0^\dagger\in {\mathfrak a}_{[11]}^{[2]},\quad a_1^\dagger\in {\mathfrak a}_{[10]}^{[2]},\quad a_2^\dagger\in {\mathfrak a}_{[01]}^{[2]}.
\eea
The Hilbert space  ${\cal H}_p^{(M)}$ is spanned by the normalized vectors $|n;r_0r_1r_2\rangle_p^{(M)}$ such that
\bea\label{spanningn4m}
|n;r_0r_1r_2\rangle_p^{(M)}&\propto& {\widehat{\Delta_p^{(M-1)}}((B_4^\dagger)^n(a_0^\dagger)^{r_0}(a_1^\dagger)^{r_1}(a_2^\dagger)^{r_2})}\cdot |0;000\rangle^{(M)},
\eea 
with $n\in {\mathbb N}_0$, while the restrictions on  $r_0,r_1,r_2$ are given below. The hat on the coproduct symbol
means that it is evaluated in the representation given by formulas (\ref{b4b4dagger}) and (\ref{aadagger}).\par
The constraints on $r_0,r_1,r_2$ are recovered by applying the same reasonings discussed in Section {\bf 4} to derive formula (\ref{n2span}). We get
\bea\label{n4restrictions}
{\textrm{for}}\quad p&=&0 ~~~~~~: \quad 0\leq r_0,r_1,r_2\leq M,\nonumber\\
{\textrm{for}} \quad p&=&1,2 ~~~: \quad 0\leq r_0\leq M;\quad r_1,r_2=0,1, \quad{\textrm{together with}}\nonumber\\
{\textrm{for}}\quad p&=&0,1,2: \quad  0\leq r_0+r_1+r_2\leq M.
\eea
The grading of the vectors $|n;r_0r_1r_2\rangle_p^{(M)}$ is determined, see (\ref{algvec}), from the assignment of the
grading of the Fock state $|0;000\rangle^{(M)}$, given by
\bea
|0;000\rangle^{(M)}\in [0] \quad {\textrm{for}}\quad p=0,1, &&
|0;000\rangle^{(M)}\in [00] \quad {\textrm{for}}\quad p=2.
\eea

Since $B_4^\dagger=b^\dagger\cdot {\mathbb I}_4$, where $b^\dagger$ is the oscillator introduced in (\ref{b4b4dagger}),
the coproduct $ {\widehat {\Delta_p}} (B_4^\dagger)$ reads ${\widehat {\Delta_p}} (B_4^\dagger)=B_4^\dagger\otimes{\mathbb I}_4+{\mathbb I}_4\otimes {B_4^\dagger}= 2 b^\dagger\cdot {\mathbb I}_{16}$; more generally we have
\bea
{\widehat {\Delta_p^{(M-1)}}}((B_4^\dagger)^n)&\propto& (b^\dagger)^n\cdot {\mathbb I}_{4^M},
\eea
so that the vectors $
|n;r_0r_1r_2\rangle_p^{(M)}$ can be expressed as
\bea\label{separation}
|n;r_0r_1r_2\rangle_p^{(M)}&\propto&(b^\dagger)^{n}\cdot{\widehat{\Delta_p^{(M-1)}}((a_0^\dagger)^{r_0}(a_1^\dagger)^{r_1}(a_2^\dagger)^{r_2})}\cdot |0;000\rangle^{(M)},
\eea
therefore separating the differential part given by the powers of $b^\dagger$. At $n=0$ the operators
${\widehat{\Delta_p^{{(M-1)}}}((a_0^\dagger)^{r_0}(a_1^\dagger)^{r_1}(a_2^\dagger)^{r_2})}$  are a set of $4^M\times 4^M$ constant real matrices whose total number $d_p^{(M)}$ is derived from the restrictions (\ref{n4restrictions}).\par
The finite-dimensional vector spaces 
 ${\cal H}_{fin;p}^{(M)}\subset {\cal H}_p^{(M)}$ of dimension  $d_p^{(M)}$ are obtained by applying the
${\widehat{\Delta_p^{{(M-1)}}}((a_0^\dagger)^{r_0}(a_1^\dagger)^{r_1}(a_2^\dagger)^{r_2})}$ coproducts to the
vacuum  $|0;000\rangle^{(M)}$. The counting of $d_p^{(M)}$ goes as follows.
In the supersymmetric and ${\mathbb Z}_2\times{\mathbb Z}_2$ grading ($p=1,2$) the restrictions (\ref{n4restrictions}) give, for 
$d_1^{(M)}=d_2^{(M)}\equiv d_{susy/{\mathbb Z}_2\times{\mathbb Z}_2}^{(M)}$,
\bea\label{counting}
M+1&&{\textrm{states}}\quad {\textrm{from}}\quad r_1=r_2=0;\nonumber\\
 2\times M &&{\textrm{states}}\quad {\textrm{from}}\quad r_1=1,~ r_2=0\quad{\textrm{and}}\quad r_1=0,~r_2=1;\nonumber\\
M-1&&{\textrm{states}}\quad {\textrm{from}}\quad r_1=r_2=1.
\eea
Therefore, the dimension $d_{susy/{\mathbb Z}_2\times{\mathbb Z}_2}^{(M)}$ of the supersymmetric and ${\mathbb Z}_2\times{\mathbb Z}_2$-graded finite  Hilbert spaces ${\cal H}_{fin;1}^{(M)}, {\cal H}_{fin;2}^{(M)}$ is
\bea
d_{susy/{\mathbb Z}_2\times{\mathbb Z}_2}^{(M)}&=& M+1+2\times M + M-1 = 4 M.
\eea

In the bosonic case the counting takes into account that, for an admissible $j>1$ given by $r_1+r_2=j$, there are $j+1$ combinations ($r_1=j, r_2=0; ~r_1=j-1,r_2=1;~\ldots ;~ r_1=0,r_2=j$) producing the same number of states. It is easily shown, after some combinatorics, that $ d_{0}^{(M)}\equiv d_{bos}^{(M)}$ is recovered from the tetrahedral numbers $T(M)$, defined as
\bea
T(M) &=& \sum_{j=0}^M P(j)\quad {\textrm{for}}\quad 
P(j) = \frac{1}{2}(j+2)(j+1).
\eea
In the above formulas $P(j)$ represents the partition of $j$ identical objects in $3$ boxes.\par
Starting from $M=0,1,2,\ldots$, the tetrahedral numbers produce the sequence \par $1,4,10,20,35,56,\ldots$ (sequence $A000292$ in the OEIS database).\par
The explicit formula for the dimension $d_{bos}^{(M)}$ of the bosonic finite Hilbert spaces
${\cal H}_0^{(M)}$ is 
\bea
d_{bos}^{(M)}&=&T(M) = \frac{1}{6}(M^3+6M^2+11M+6).
\eea\par
~\par
{\it Remark}: In the tensor product construction, the $M$-particle  ${\cal N}=4$  Hamiltonians are realized by 
$4^M\times 4^M$ matrices with differential entries.  Since, however, the $p=0,1,2$ physical Hilbert spaces ${\cal H}_p^{(M)}$ are, see (\ref{subset}), a subset of $({\cal H}^{(1)})^{\otimes M}$,  the Hamiltonians can be projected onto  the ${\cal H}_p^{(M)}$ subspaces and therefore expressed by $
d_p^{(M)}\times d_p^{(M)}$ matrices with differential entries. In the supersymmetric and ${\mathbb Z}_2\times{\mathbb Z}_2$-graded cases the vectors which respectively belong to ${\cal H}_1^{(M)}, {\cal H}_2^{(M)}$ have
$d_{susy/{\mathbb Z}_2\times{\mathbb Z}_2}^{(M)}$ components, while in the bosonic case the vectors in ${\cal H}_1^{(M)}$ have $d_{bos}^{(M)}$ components. \par
It has to be noted, in particular,  that while the size of the columns/rows of $({\cal H}^{(1)})^{\otimes M}$
grows exponentially with $M$ as $4^M$ ($4,16,64,256,\ldots $), the dimension  of the supersymmetric and ${\mathbb Z}_2\times{\mathbb Z}_2$-graded column vectors grows linearly as $4M$
($4,8,12,16,\ldots$). This counting allows to make the connection between the present construction and the derivation of the $2$-particle ${\mathbb Z}_2\times {\mathbb Z}_2$-graded invariant quantum Hamiltonians obtained in \cite{aktquant} by quantizing classical actions. In that paper the Hamiltonians, the observables and the creation operators are $8\times 8$ matrix differential operators. The ${\cal N}=4$ oscillator Hamiltonian under consideration here is a specfic example of the large class of ${\mathbb Z}_2\times {\mathbb Z}_2$-graded invariant Hamiltonians (which even allow for the presence of multiparticle interacting terms) obtained in \cite{aktquant}.

\subsection{Energy degeneracies and quantum numbers}

The first set of compatible observables $H_4, N_f, N_v, N_w$ introduced in (\ref{h4nf}) and (\ref{nvnw}) induces an abelian algebra ${\mathfrak{a}}$. Let ${\cal U}\equiv {\cal U}({\mathfrak{a}})$ be its associated Universal Enveloping Algebra and let  $g\in{{\mathfrak{a}}}$ denote one of the four observables mentioned above. The $M$-particle
observable induced by $g$  is recovered by evaluating the coproduct $\Delta^{(m-1)}(g)\in {\cal U}\otimes \ldots \otimes {\cal U}$ in the representation given by (\ref{h4nf},\ref{nvnw}). The corresponding operator
${\widehat{\Delta^{(M-1)}}}(g)$ will be denoted, for simplicity, as $g^{(M)}$.\par
The four derived operators $H_4^{(M)}, ~N_f^{(M)},N_v^{(M)},N_w^{(M)}$ define a complete set of compatible observables
for each one of the $p=0,1,2$ variants of the $M$-particle Hilbert space ${\cal H}_p^{(M)}$. Their common eigenvectors $|n;r_0r_1r_2\rangle_p^{(M)}$ have been introduced in (\ref{spanningn4m}). Their respective
eigenvalues are
\bea\label{n4multiparteigen}
H_4^{(M)}|n;r_0r_1r_2\rangle_p^{(M)}&=&(n+r_1+r_2)|n;r_0r_1r_2\rangle_p^{(M)},\nonumber\\
N_f^{(M)}|n;r_0r_1r_2\rangle_p^{(M)}&=&(r_1+r_2)|n;r_0r_1r_2\rangle_p^{(M)},\nonumber\\
N_v^{(M)}|n;r_0r_1r_2\rangle_p^{(M)}&=&(r_0+r_2)|n;r_0r_1r_2\rangle_p^{(M)},\nonumber\\
N_w^{(M)}|n;r_0r_1r_2\rangle_p^{(M)}&=&(r_0+r_1)|n;r_0r_1r_2\rangle_p^{(M)}.
\eea
The spectrum of the multiparticle Hamiltonian $H_4^{(M)}$ is given by the set of non-negative integers $E=0,1,2,3,\ldots$. The degeneracy of each energy level $E$ is computed by taking into account the
(\ref{n4restrictions}) restrictions on $r_0,r_1,r_2$.\par
In the supersymmetric and ${\mathbb Z}_2\times{\mathbb Z}_2$-graded cases (which will be denoted with the subscript $p=1,2$) we obtain, adapting formula (\ref{counting}), the degeneracy of the energy level $E$. To  avoid confusion with other symbols, we indicate the degeneracy in this Section as ``$\sharp_{1,2}(E)$". We get
\bea
\sharp_{1,2}^{(M)}(0)&=&M+1, ~({\textrm{states obtained from}}~ n=r_1=r_2=0);\nonumber\\
\sharp_{1,2}^{(M)}(1)&=&3M+1, ~ (M+1~ {\textrm{states from}}~n=1, r_1=r_2=0; ~2M~{\textrm{from}}~
n=0, r_1+r_2=1);\nonumber\\
\sharp_{1,2}^{(M)}(2)&=&4M, \quad(3M+1~{\textrm{states from}}~n=1,2; ~M-1 ~{\textrm{states from}}~ n=0, r_1+r_2=2).\nonumber\\&&
\eea
The $E>2$ counting repeats the above $E=2$ counting with the shift, in the right hand side, $n\rightarrow n+E-2$, so that
\bea
&\sharp_{1,2}^{(M)}(0) = M+1,\qquad
\sharp_{1,2}^{(M)}(1) = 3M+1,\qquad
\sharp_{1,2}^{(M)}(E) = 4M \quad{\textrm{for}} \quad E\geq 2.&
\eea
The counting in the ($p=0$) bosonic case of the  $\sharp_0^{(M)}(E)$ degeneracy for $E\in {\mathbb N}_0$, $M\in {\mathbb N}$ is based on the expansion
\bea 
& (M+1) +2M + 3 (M-1)+\ldots+ (j+1)(M-j+1)+\ldots = \sum_{j=0}^E (j+1)(M-j+1).\qquad&
\eea
The sum of the above series gives two terms, one linear in $M$ and the other one depending only on $E$. We have
\bea
\sum_{j=0}^E (j+1)(M-j+1)&=& A(E)\cdot M-B(E),\qquad{\textrm{with}} \nonumber\\
\quad A(E)=\frac{1}{2}(E^2+3E+2), && B(E)= \frac{1}{6}(2E-3)(E+2)(E+1).
\eea
Statring from $E=0,1,2,\ldots$, the term $B(E)$ produces the sequence $-1,-1,2,10,25,49,\ldots$ (sequence 
$A058373$ in the OEIS database).\par
We get, for the bosonic degeneracy $\sharp_{0}^{(M)}(E)$,
\bea
\sharp_{0}^{(M)}(E)&=&A(E)M-B(E)\qquad\qquad\qquad\qquad \quad ~~~~{\textrm{for}} \quad M\leq E,\nonumber\\
\sharp_{0}^{(M)}(E)&=& \sharp_{0}^{(E)}(E)=\frac{1}{6}(E^3+6E^2+11E+6)\qquad ~ {\textrm{for}} \quad M> E.
\eea 
The energy levels degeneracy grows linearly with $M$ until reaching the maximal value at $M=E$.\par
The following table is useful in order to compare, for low energy values $E$ and low particle numbers $M$, the degeneracy of the bosonic (denoted with the ``$B$" subscript) versus the supersymmetric/${\mathbb Z}_2\times {\mathbb Z}_2$-graded (denoted with ``$S$") cases. We have
\bea\label{bsdegeneracy}&
\begin{array}{|c||c|c||c|c||c|c||c|c||c|c||c|c||}\hline M: &1_B&1_S&2_B&2_S&3_B&3_S&4_B&4_S&5_B&5_S\\ \hline \hline
E=0:&2&2&3&3&4&4&5&5&6&6\\ \hline 
E=1:&4&4&7&7&10&10&13&13&16&16\\ \hline
E=2:&4&4&10&8&16&12&22&16&28&20\\ \hline 
E=3:&4&4&10&8&20&12&30&16&40&20\\ \hline 
E=4:&4&4&10&8&20&12&35&16&50&20\\ \hline 
E=5:&4&4&10&8&20&12&35&16&56&20\\ \hline 
\end{array}&
\eea\\
~\\
{\it Comment 1}: in all three variants of the theory the vacuum state is $(M+1)$-degenerate and the first excited level is $(3M+1)$-degenerate; furthermore, these states present the same quantum numbers.\par
~\\
{\it Comment 2}: starting from the energy level $E=2$ the bosonic variant is discriminated with respect to the supersymmetric/${\mathbb Z}_2\times {\mathbb Z}_2$-graded variants of the theory for the presence of extra states characterized by different quantum numbers. At $E=2$, e.g.,  two extra states are found. They correspond to
$|0;020\rangle_0^{(M)}$ and $
|0;002\rangle_0^{(M)}$; their respective sets of $(H_4^{(M)}, ~N_f^{(M)},N_v^{(M)},N_w^{(M)})$ eigenvalues are
$(2,2,0,2)$ and $(2,2,2,0)$. In the supersymmetric/${\mathbb Z}_2\times {\mathbb Z}_2$-graded variants the states with these quantum numbers are absent due to the Pauli exclusion principle for fermions.
\par
~\\
{\it Comment 3}: so far, with measurements conducted with the observables $H_4^{(M)}, ~N_f^{(M)},~N_v^{(M)},N_w^{(M)}$, the supersymmetric Hilbert space ${\cal H}_1^{(M)}$ and the ${\mathbb Z}_2\times{\mathbb Z}_2$-graded Hilbert space ${\cal H}_2^{(M)}$ cannot be discriminated. They lead to equivalent theories {\it as far as measurements which only involve  $H_4^{(M)}, ~N_f^{(M)},~N_v^{(M)},N_w^{(M)}$ are performed}.

\subsection{The $M=2,3$-particle Hilbert spaces}

We present here, for later convenience, the explicit formulas of the $2$-particle states belonging to the Hilbert spaces
${\cal H}_p^{(2)}$ for all three $p=0,1,2$ variants of the theory and of the $3$-particle states belonging to ${\cal H}_p^{(3)}$ for the $p=1,2$ supersymmetric/${\mathbb Z}_2\times{\mathbb Z}_2$-graded variants.\par
For any $p=0,1,2$, the $2$-particle $H_4^{(2)}$ and $3$-particle $H_4^{(3)}$ Hamiltonians can be expressed as
\bea
H_4^{(2)}&=&\frac{1}{2}(-\partial_{x_1}^2-\partial_{x_2}^2+x_1^2+x_2^2-2)\cdot {\mathbb I}_8 + N_f^{(2)}\quad{\textrm{with}}\quad N_f^{(2)}={\mathbb I}_4\otimes N_f+N_f\otimes {\mathbb I}_4,\nonumber\\
H_4^{(3)}&=&\frac{1}{2}(-\partial_{x_1}^2-\partial_{x_2}^2-\partial_{x_3}^2+x_1^2+x_2^2+x_3^2-3)\cdot {\mathbb I}_{64} + N_f^{(3)} \quad {\textrm{with}}\quad\nonumber\\
&& N_f^{(3)}={\mathbb I}_4\otimes {\mathbb I}_4\otimes N_f+{\mathbb I}_4\otimes N_f\otimes {\mathbb I}_4+N_f\otimes{\mathbb I}_4\otimes{\mathbb I}_4.
\eea
The diagonal operator $N_f$ was introduced in (\ref{h4nf}). \par
The states will be written in terms of the column vectors $v_i$, with entry $1$ in the $i$-th position and $0$ otherwise. In the $2$-particle case the vectors have $16$ components and $i=1,2,\ldots,16$ while, in the $3$-particle case, we have $i=1,2,\ldots, 64$ for the $64$-component vectors.
\par
Since, see formula (\ref{separation}), the $
|n;r_0r_1r_2\rangle_p^{(M)}$ states are easily obtained by applying the creation operator $(b^\dagger)^n$ to
the $
|0;r_0r_1r_2\rangle_p^{(M)}$ states, it is sufficient for our purposes to present the $p=0,1,2$ finite bases of $n=0$ orthonormal vectors. We have indeed
\bea\label{finitefull}
|0;r_0r_1r_2\rangle_p^{(M)}\in {\cal H}_{fin;p}^{(M)} &\Rightarrow |n;r_0r_1r_2\rangle_p^{(M)}=\frac{1}{\sqrt{n!}}(b^\dagger)^n|0;r_0r_1r_2\rangle_p^{(M)}\in{\cal H}_{p}^{(M)},
\eea
with $b^\dagger$ given in (\ref{b4b4dagger}).\par
The $M$-particle states are given in terms of tensor products of the normalized single-particle vectors $|0;\delta_0\delta_1\delta_2\rangle$
(for
$ 0\leq\delta_0+\delta_1+\delta_2\leq 1$) given in (\ref{singlen4vectors}). \par
The results are summarized as follows:
\newpage 
{\it I - Bosonic $2$-particle case}:\par
~\par
In the $2$-particle case we have that the finite-dimensional bosonic Hilbert space ${\cal H}_{fin;0}^{(2)}$ is spanned by the $10$ orthonormal vectors
\bea
|0;000\rangle_0^{(2)}&=&{\footnotesize{|0;000\rangle\otimes |0;000\rangle= G_2\cdot v_1}},\nonumber\\
|0;100\rangle_0^{(2)}&=&{\footnotesize{\frac{1}{\sqrt{2}}\Big(|0;100\rangle\otimes |0;000\rangle+|0;000\rangle\otimes |0;100\rangle\Big)=\frac{G_2}{\sqrt{2}}(v_2+v_5)}},\nonumber\\
|0;010\rangle_0^{(2)}&=&{\footnotesize{\frac{1}{\sqrt{2}}\Big(|0;010\rangle\otimes |0;000\rangle+|0;000\rangle\otimes |0;010\rangle\Big)=\frac{G_2}{\sqrt{2}}(v_3+v_9)}},\nonumber\\
|0;001\rangle_0^{(2)}&=&{\footnotesize{\frac{1}{\sqrt{2}}\Big(|0;001\rangle\otimes |0;000\rangle+|0;000\rangle\otimes |0;001\rangle\Big)=\frac{G_2}{\sqrt{2}}(v_4+v_{13})}},\nonumber\\
|0;110\rangle_0^{(2)}&=&{\footnotesize{\frac{1}{\sqrt{2}}(|0;100\rangle\otimes |0;010\rangle+|0;010\rangle\otimes |0;100\rangle)=\frac{G_2}{\sqrt{2}}(v_7+v_{10})}},\nonumber\\
|0;101\rangle_0^{(2)}&=&{\footnotesize{\frac{1}{\sqrt{2}}\Big(|0;100\rangle\otimes |0;001\rangle+|0;001\rangle\otimes |0;100\rangle\Big)=\frac{G_2}{\sqrt{2}}(v_8+v_{14})}},\nonumber\\
|0;011\rangle_0^{(2)}&=&{\footnotesize{\frac{1}{\sqrt{2}}\Big(|0;010\rangle\otimes |0;001\rangle+|0;001\rangle\otimes |0;010\rangle\Big)=\frac{G_2}{\sqrt{2}}(v_{12}+v_{15})}},\nonumber\\
|0;200\rangle_0^{(2)}&=&{\footnotesize{|0;100\rangle\otimes |0;100\rangle= G_2\cdot v_6}},\nonumber\\
|0;020\rangle_0^{(2)}&=&{\footnotesize{|0;010\rangle\otimes |0;010\rangle= G_2\cdot v_{11}}},\nonumber\\
|0;002\rangle_0^{(2)}&=&{\footnotesize{|0;001\rangle\otimes |0;001\rangle= G_2\cdot v_{16}}}.
\eea
The factor $G_2$ is the Gaussian term $G_2={\footnotesize{\frac{1}{\sqrt{\pi}}e^{-\frac{1}{2}(x_1^2+x_2^2)}}}$.
\newpage
{\it II - Supersymmetric and ${\mathbb Z}_2\times {\mathbb Z}_2$-graded $2$-particle cases}:\par
~\par
It is convenient to introduce a unified framework to express the ${\mathbb Z}_2^p$-graded Hilbert spaces ${\cal H}_{fin;p}^{(M)}$ for  $p=1,2$  by introducing a
 $\varepsilon = (-1)^p$ sign ($\varepsilon =-1$ for the supersymmetric case, $\varepsilon=1$ for the ${\mathbb Z}_2\times{\mathbb Z}_2$-graded case). The vectors spanning ${\cal H}_{fin;p}^{(2)}$ for
$p= \frac{1}{2}(3+\varepsilon)$ will be denoted as
$|0;r_0r_1r_2\rangle_{\frac{1}{2}(3+\varepsilon)}^{(2)}\equiv|0;r_0r_1r_2\rangle_{[\varepsilon]}^{(2)} $. \par
The $8$ orthonormal vectors of the respective $\varepsilon=\pm 1$ Hilbert spaces are
\bea\label{2psusyz2z2}
|0;000\rangle_{[\varepsilon]}^{(2)}&=&{\footnotesize{|0;000\rangle\otimes |0;000\rangle= G_2\cdot v_1}},\nonumber\\
|0;100\rangle_{[\varepsilon]}^{(2)}&=&{\footnotesize{\frac{1}{\sqrt{2}}\Big(|0;100\rangle\otimes |0;000\rangle+|0;000\rangle\otimes |0;100\rangle\Big)=\frac{G_2}{\sqrt{2}}(v_2+v_5)}},\nonumber\\
|0;010\rangle_{[\varepsilon]}^{(2)}&=&{\footnotesize{\frac{1}{\sqrt{2}}\Big(|0;010\rangle\otimes |0;000\rangle+|0;000\rangle\otimes |0;010\rangle\Big)=\frac{G_2}{\sqrt{2}}(v_3+v_9)}},\nonumber\\
|0;001\rangle_{[\varepsilon]}^{(2)}&=&{\footnotesize{\frac{1}{\sqrt{2}}\Big(|0;001\rangle\otimes |0;000\rangle+|0;000\rangle\otimes |0;001\rangle\Big)=\frac{G_2}{\sqrt{2}}(v_4+v_{13})}},\nonumber\\
|0;200\rangle_{[\varepsilon]}^{(2)}&=&{\footnotesize{|0;100\rangle\otimes |0;100\rangle= G_2\cdot v_6}},\nonumber\\
|0;110\rangle_{[\varepsilon]}^{(2)}&=&{\footnotesize{\frac{1}{\sqrt{2}}\Big(|0;100\rangle\otimes |0;010\rangle-\varepsilon |0;010\rangle\otimes |0;100\rangle\Big)=\frac{G_2}{\sqrt{2}}(v_7-\varepsilon v_{10})}},\nonumber\\
|0;101\rangle_{[\varepsilon]}^{(2)}&=&{\footnotesize{\frac{1}{\sqrt{2}}\Big(|0;100\rangle\otimes |0;001\rangle-\varepsilon |0;001\rangle\otimes |0;100\rangle\Big)=\frac{G_2}{\sqrt{2}}(v_8-\varepsilon v_{14})}},\nonumber\\
|0;011\rangle_{[\varepsilon]}^{(2)}&=&{\footnotesize{\frac{1}{\sqrt{2}}\Big(|0;010\rangle\otimes |0;001\rangle+\varepsilon |0;001\rangle\otimes |0;010\rangle\Big)=\frac{G_2}{\sqrt{2}}(v_{12}+\varepsilon v_{15})}}.
\eea
Some comments are in order:\\
~\\
{\it i}) the supersymmetric and ${\mathbb Z}_2\times {\mathbb Z}_2$-graded Hilbert spaces are not subspaces of the bosonic Hilbert space (${\cal H}_{fin;1,2}^{(2)}  \nsubset {\cal H}_{fin;0}^{(2)}$) since, e.g., in the supersymmetric case $|0;011\rangle_{[-1]}^{(2)}\not\in {\cal H}_{fin;0}^{(2)}$ and, in the ${\mathbb Z}_2\times {\mathbb Z}_2$-graded case, $
|0;110\rangle_{[+1]}^{(2)},
|0;101\rangle_{[+1]}^{(2)}\not\in {\cal H}_{fin;0}^{(2)}$;\\
~\\
{\it ii}) as vector spaces, the intersection ${\cal H}_{fin;1}^{(2)}\cap {\cal H}_{fin;2}^{(2)}\neq \emptyset$ is spanned by the first $5$ vectors in formula (\ref{2psusyz2z2}). These vectors do not depend on the $\varepsilon $ sign. Since, however, the last three vectors differ due to the presence of $\varepsilon$ in their right hand side, then
${\cal H}_{fin;1}^{(2)}\neq {\cal H}_{fin;2}^{(2)}$. From (\ref{finitefull}) we get ${\cal H}_{1}^{(2)}\neq {\cal H}_{2}^{(2)}$;\\
~\\
{\it iii}) the (\ref{2psusyz2z2}) states are eigenvectors with $+1$ eigenvalue of the $S_{12}^{(2)}$ permutation operator introduced in ({\ref{S12}). $S_{12}^{(2)}$ exchanges the first and second spaces in the right hand side tensor products; the (anti)symmetry of the wave functions is automatically taken into account by the signs obtained from braiding the tensors.
\newpage

{\it III - Supersymmetric and ${\mathbb Z}_2\times {\mathbb Z}_2$-graded $3$-particle cases}:\par
~\par
We consider here the $3$-particle cases with $p=1,2$. In the $\varepsilon$-unified notation, the $12$ orthonormal vectors spanning
for $\varepsilon=\pm 1$ the respective
${\cal H}_{fin;\frac{1}{2}(3+\varepsilon)}^{(3)}$ Hilbert spaces are
\bea\label{3psusyz2z2}
|0;000\rangle_{[\varepsilon]}^{(3)}&=&{\footnotesize{|0;000\rangle\otimes |0;000\rangle\otimes |0;000\rangle = G_3\cdot v_1,}}\nonumber\\
|0;100\rangle_{[\varepsilon]}^{(3)}&=&{\footnotesize{\frac{1}{\sqrt{3}}\Big(|0;100\rangle\otimes |0;000\rangle\otimes |0;000\rangle+
|0;000\rangle\otimes |0;100\rangle\otimes |0;000\rangle +}}\nonumber\\
&&{\footnotesize{|0;000\rangle\otimes |0;000\rangle\otimes |0;100\rangle  \Big)= \frac{G_3}{\sqrt{3}}\cdot (v_2+v_5+v_{17}),}}\nonumber\\
|0;010\rangle_{[\varepsilon]}^{(3)}&=& {\footnotesize{\frac{1}{\sqrt{3}}\Big(|0;010\rangle\otimes |0;000\rangle\otimes |0;000\rangle+
|0;000\rangle\otimes |0;010\rangle\otimes |0;000\rangle +}}\nonumber\\
&&{\footnotesize{|0;000\rangle\otimes |0;000\rangle\otimes |0;010\rangle  \Big)= \frac{G_3}{\sqrt{3}}\cdot (v_3+v_{9}+v_{33}),}}    \nonumber\\
|0;001\rangle_{[\varepsilon]}^{(3)}&=&{\footnotesize{\frac{1}{\sqrt{3}}\Big(|0;001\rangle\otimes |0;000\rangle\otimes |0;000\rangle+
|0;000\rangle\otimes |0;001\rangle\otimes |0;000\rangle +}}\nonumber\\
&&{\footnotesize{|0;000\rangle\otimes |0;000\rangle\otimes |0;001\rangle  \Big)= \frac{G_3}{\sqrt{3}}\cdot (v_4+v_{13}+v_{49}),}}\nonumber\\
|0;200\rangle_{[\varepsilon]}^{(3)}&=&
{\footnotesize{\frac{1}{\sqrt{3}}\Big(|0;100\rangle\otimes |0;100\rangle\otimes |0;000\rangle+
|0;100\rangle\otimes |0;000\rangle\otimes |0;100\rangle +}}\nonumber\\
&&{\footnotesize{|0;000\rangle\otimes |0;100\rangle\otimes |0;100\rangle  \Big)= \frac{G_3}{\sqrt{3}}\cdot (v_6+v_{18}+v_{21}),}}
\nonumber\\
|0;300\rangle_{[\varepsilon]}^{(3)}&=&{\footnotesize{|0;100\rangle\otimes |0;100\rangle\otimes |0;100\rangle = G_3\cdot v_{22},}}\nonumber\\
|0;110\rangle_{[\varepsilon]}^{(3)}&=&
{\footnotesize{\frac{1}{\sqrt{6}}\Big(|0;000\rangle\otimes |0;100\rangle\otimes |0;010\rangle-\varepsilon
|0;000\rangle\otimes |0;010\rangle\otimes |0;100\rangle +}}\nonumber\\
&&{\footnotesize{|0;100\rangle\otimes |0;000\rangle\otimes |0;010\rangle -\varepsilon|0;010\rangle\otimes |0;000\rangle\otimes |0;100\rangle +}}\nonumber\\
&&{\footnotesize{|0;100\rangle\otimes |0;010\rangle\otimes |0;000\rangle -\varepsilon|0;010\rangle\otimes |0;100\rangle\otimes |0;000\rangle \Big)}}=\nonumber\\
&&{\footnotesize{ \frac{G_3}{\sqrt{6}}\cdot (v_7-\varepsilon v_{10}+v_{19}+v_{25}-\varepsilon v_{34} -\varepsilon v_{37}),}}
\nonumber\\
|0;101\rangle_{[\varepsilon]}^{(3)}&=&
{\footnotesize{\frac{1}{\sqrt{6}}\Big(|0;000\rangle\otimes |0;100\rangle\otimes |0;001\rangle-
\varepsilon|0;000\rangle\otimes |0;001\rangle\otimes |0;100\rangle +}}\nonumber\\
&&{\footnotesize{|0;100\rangle\otimes |0;000\rangle\otimes |0;001\rangle -\varepsilon|0;001\rangle\otimes |0;000\rangle\otimes |0;100\rangle +}}\nonumber\\
&&{\footnotesize{|0;100\rangle\otimes |0;001\rangle\otimes |0;000\rangle -\varepsilon|0;001\rangle\otimes |0;100\rangle\otimes |0;000\rangle \Big)}}=\nonumber\\
&&{\footnotesize{ \frac{G_3}{\sqrt{6}}\cdot (v_8-\varepsilon v_{14}+v_{20}+v_{29}-\varepsilon v_{50} -\varepsilon v_{53}),}}\nonumber\\
|0;011\rangle_{[\varepsilon]}^{(3)}&=&
{\footnotesize{\frac{1}{\sqrt{6}}\Big(|0;000\rangle\otimes |0;100\rangle\otimes |0;001\rangle+
\varepsilon|0;000\rangle\otimes |0;001\rangle\otimes |0;010\rangle +}}\nonumber\\
&&{\footnotesize{|0;010\rangle\otimes |0;000\rangle\otimes |0;001\rangle +\varepsilon|0;001\rangle\otimes |0;000\rangle\otimes |0;010\rangle +}}\nonumber\\
&&{\footnotesize{|0;010\rangle\otimes |0;001\rangle\otimes |0;000\rangle +\varepsilon|0;001\rangle\otimes |0;010\rangle\otimes |0;000\rangle \Big)}}=\nonumber\\
&&{\footnotesize{ \frac{G_3}{\sqrt{6}}\cdot (v_{12}+\varepsilon v_{15}+v_{36}+v_{45}+\varepsilon v_{51} +\varepsilon v_{57}),}}\nonumber
\eea
\bea
|0;111\rangle_{[\varepsilon]}^{(3)}&=&
{\footnotesize{\frac{1}{\sqrt{6}}\Big(|0;100\rangle\otimes |0;010\rangle\otimes |0;001\rangle+\varepsilon
|0;100\rangle\otimes |0;001\rangle\otimes |0;010\rangle -}}\nonumber\\
&&{\footnotesize{\varepsilon |0;010\rangle\otimes |0;100\rangle\otimes |0;001\rangle -|0;001\rangle\otimes |0;100\rangle\otimes |0;010\rangle +}}\nonumber\\
&&{\footnotesize{|0;010\rangle\otimes |0;001\rangle\otimes |0;100\rangle +\varepsilon|0;001\rangle\otimes |0;010\rangle\otimes |0;100\rangle \Big)}}=\nonumber\\
&&{\footnotesize{ \frac{G_3}{\sqrt{6}}\cdot (v_7-\varepsilon v_{10}+v_{19}+v_{25}-\varepsilon v_{34} -\varepsilon v_{37}),}}\nonumber\\
|0;210\rangle_{[\varepsilon]}^{(3)}&=&
{\footnotesize{\frac{1}{\sqrt{3}}\Big(|0;100\rangle\otimes |0;100\rangle\otimes |0;010\rangle-
\varepsilon |0;100\rangle\otimes |0;010\rangle\otimes |0;100\rangle +}}\nonumber\\
&&{\footnotesize{|0;010\rangle\otimes |0;100\rangle\otimes |0;100\rangle  \Big)= \frac{G_3}{\sqrt{3}}\cdot (v_{23}-\varepsilon v_{26}+v_{38}),}}\nonumber\\
|0;201\rangle_{[\varepsilon]}^{(3)}&=&
{\footnotesize{\frac{1}{\sqrt{3}}\Big(|0;100\rangle\otimes |0;100\rangle\otimes |0;001\rangle-\varepsilon |0;100\rangle\otimes |0;001\rangle\otimes |0;100\rangle +}}\nonumber\\
&&{\footnotesize{|0;001\rangle\otimes |0;100\rangle\otimes |0;100\rangle  \Big)= \frac{G_3}{\sqrt{3}}\cdot (v_{24}-\varepsilon v_{30}+v_{54}).}}
\eea
The factor $G_3$ is the Gaussian term $G_3={\footnotesize {\pi^{-\frac{3}{4}} e^{-\frac{1}{2}(x_1^2+x_2^2+x_3^2)}}}$.\par
~\par
Analogous properties as those encountered in the $2$-particle case hold.
The first $6$ vectors in formula (\ref{3psusyz2z2}) do not depend on the $\varepsilon $ sign, while the remaining $6$ vectors differ due to the presence of $\varepsilon$ in their right hand side. Therefore ${\cal H}_{fin;1}^{(3)}\cap {\cal H}_{fin;2}^{(3)}\neq \emptyset$ and
${\cal H}_{fin;1}^{(3)}\neq {\cal H}_{fin;2}^{(3)}$.\par
The (\ref{3psusyz2z2}) states are eigenvectors with $+1$ eigenvalue of the $S_{12}^{(3)}, S_{23}^{(3)}$ permutation operators introduced in ({\ref{S12S13}). These operators generate the ${\bf S}_3$ group of permutations of the three spaces appearing in the tensor products.\par
~\par
The extension of the $M=2,3$ construction to the $M>3$ particle cases is immediate.

\section{Discriminating ${\mathbb Z}_2$-graded versus ${\mathbb Z}_2\times{\mathbb Z}_2$-graded oscillators}

The core results of this paper are presented in this Section.\par
In the multiparticle sector the bosonic variant of the ${\cal N}=4$ oscillator is easily discriminated with respect to the
supersymmetric and ${\mathbb Z}_2\times {\mathbb Z}_2$-graded variants for the presence of extra states with different quantum numbers. This is reflected in different degeneracies of the energy levels, see e.g. table (\ref{bsdegeneracy}), of the bosonic case with respect to the two other cases. \par
The open question is whether the supersymmetric and ${\mathbb Z}_2\times{\mathbb Z}_2$-graded variants can be differentiated. The positive answer is given here. We recall from Section {\bf 5} that the $M>1$ multiparticle supersymmetric ${\cal H}_1^{(M)}$ and ${\mathbb Z}_2\times {\mathbb Z}_2$-graded ${\cal H}_2^{(M)}$ Hilbert spaces differ, see e.g. the comment {\it ii} concerning the states presented in formula (\ref{2psusyz2z2}); on the other hand we noted, see  comment $3$ of subsection {\bf 5.3}, that these Hilbert  spaces cannot be discriminated with measurements conducted with multiparticle observables induced by the first set ($H_4,N_f,N_v,N_w$) of $4$ complete single-particle observables.\par
The key ingredient to differentiate, for $M>1$, the Hilbert space ${\cal H}_2^{(M)}$ from ${\cal H}_1^{(M)}$ is to conduct measurements of multiparticle observables induced by the second set of $4$ complete single-particle observables, given by the operators $H_4,N_f,N_T,X_f$ introduced in (\ref{h4nf},\ref{nvnw},\ref{xf}). The multiparticle operators induced by the Fermion Exchange Operator $X_f$ allow to make the distinction. Let's see how this happens.\par
To prove our point is sufficient to consider just $M=2$. In any case, explicit formulas will also be given for $M=3$ (the extension to $M>3$ particles is immediate).\par
The operators $H_4,N_f,N_T,X_f$ induce an abelian (all commutators are vanishing) algebra ${\overline{\mathfrak a}}$. As usual (see Appendix {\bf B}), ${\cal U}\equiv{\cal U}({\overline{\mathfrak a}})$ denotes its Universal Enveloping Algebra; the coproduct of a primitive element, for $g\in{\overline{\mathfrak {a}}}$, is denoted as
$\Delta^{(M-1)}(g)\in {\cal U}^{\otimes M}$, while ${\widehat {\Delta^{(M-1)}(g)}}$ is the evaluation of the coproduct in the representation of ${\overline{\mathfrak{a}}}$  provided by formulas (\ref{h4nf},\ref{nvnw},\ref{xf}). \par 
Besides the operators $g^{(M)}\equiv {\widehat {\Delta^{(M-1)}(g)}}$, the considerations below also  involve the operators ${\widehat{{\Delta^{(M-1)}(g^2)}}}= {\widehat{{\Delta^{(M-1)}(g)}}}\cdot {\widehat{{\Delta^{(M-1)}(g)}}}$ for $g=X_f$.\par
Overall, for the $2$-particle case the list of operators that we are discussing is
\bea\label{2psecondsetop}
H_4^{(2)}=H_4\otimes{\mathbb I}_4+{\mathbb I}_4\otimes H_4,  &&
N_f^{(2)}=N_f\otimes{\mathbb I}_4+{\mathbb I}_4\otimes N_f,\nonumber\\
N_T^{(2)}=N_T\otimes{\mathbb I}_4+{\mathbb I}_4\otimes N_T, &&
X_f^{(2)}=X_f\otimes{\mathbb I}_4+{\mathbb I}_4\otimes X_f, 
\eea
together with
\bea\label{2psecondsetop2}
& \big(X_f^{(2)}\cdot X_f^{(2)} \big):= Z_{2f}= Y_{2f}+{W}_{2f},\quad{\textrm{for}}\quad
Y_{2f}=(X_f)^2\otimes{\mathbb I}_4+{\mathbb I}_4\otimes (X_f)^2,\quad {W}_{2f} = 2 X_f\otimes X_f.&\nonumber\\
\eea
For the $3$-particle case the list is
\bea\label{3psecondsetop}
H_4^{(3)}&=&H_4\otimes{\mathbb I}_4\otimes{\mathbb I}_4+{\mathbb I}_4\otimes H_4\otimes{\mathbb I}_4+{\mathbb I}_4\otimes {\mathbb I}_4\otimes{ H}_4, \nonumber\\
N_f^{(3)}&=&N_f\otimes{\mathbb I}_4\otimes{\mathbb I}_4+{\mathbb I}_4\otimes N_f\otimes{\mathbb I}_4+{\mathbb I}_4\otimes {\mathbb I}_4\otimes N_f, \nonumber\\
N_T^{(3)}&=&N_T\otimes{\mathbb I}_4\otimes{\mathbb I}_4+{\mathbb I}_4\otimes N_T\otimes{\mathbb I}_4+{\mathbb I}_4\otimes {\mathbb I}_4\otimes N_T, \nonumber\\
X_f^{(3)}&=&X_f\otimes{\mathbb I}_4\otimes{\mathbb I}_4+{\mathbb I}_4\otimes X_f\otimes{\mathbb I}_4+{\mathbb I}_4\otimes {\mathbb I}_4\otimes X_f, 
\eea
together with
\bea\label{3psecondsetop2}
\big(X_f^{(3)}\cdot X_f^{(3)} \big)&:=& Z_{3f}= Y_{3f}+{W}_{3f},\qquad {\textrm{for}}\nonumber\\
Y_{3f}\qquad&=&(X_f)^2\otimes{\mathbb I}_4\otimes{\mathbb I}_4+{\mathbb I}_4\otimes (X_f)^2\otimes{\mathbb I}_4+{\mathbb I}_4\otimes{\mathbb I}_4\otimes ( X_f)^2,\nonumber\\
{W}_{3f}\qquad &=& 2\cdot(X_f\otimes X_f\otimes{\mathbb I}_4+X_f\otimes {\mathbb I}_4\otimes X_f+{\mathbb I}_4\otimes{X}_f\otimes X_f).
\eea

All $2$-particle operators given in formulas (\ref{2psecondsetop},\ref{2psecondsetop2}) mutually commute.\par
All $3$-particle operators given in formulas (\ref{3psecondsetop},\ref{3psecondsetop2}) mutually commute.\par
In the supersymmetric case all above operators are bosonic ($0$-graded). \par
In the ${\mathbb Z}_2\times {\mathbb Z}_2$-graded case the grading assignment of the above operators is
\bea\label{0011gradingobs}
[00]&:& H_4^{(2)},N_f^{(2)},N_T^{(2)}, Y_{2f}, {W}_{2f}, Z_{2f},H_4^{(3)}, H_4^{(3)},N_f^{(3)},N_T^{(3)}, Y_{3f}, W_{3f}
,{Z}_{3f},\nonumber\\
\relax [11]&:& X_f^{(2)}, X_f^{(3)}.
\eea
The $11$-grading of $X_f^{(2)}, X_f^{(3)}$ is a consequence of the fact that the single-particle operator $X_f$ 
exchanges the two types of fermions, see (\ref{xf}) and the following ``Remark 2". The $00$-grading of $Z_{2f}, Z_{3f}$ (and the related operators $Y_{2f},W_{2f},Y_{3f},W_{3f}$) follows, by consistency, from (\ref{gradingsum}).\par
One would expect, on a general ground, that a Hermitian, observable operator in a ${\mathbb Z}_2\times {\mathbb Z}_2$-graded theory should belong to the $00$-sector since the result of a measurement should be a real eigenvalue. If the operator would  belong to a different sector, like the 11-grading, by consistency its eigenvalues should also be $11$-graded and not just real numbers.\par
This point is better illustrated if, for the time being, we disregard the gradings of the Hilbert spaces  ${\cal H}_1^{(M=2,3)},~ {\cal H}_2^{(M=2,3)}$ and just focus on the action of the matrix operators $X_f^{(2)}$, $X_f^{(3)}$. \par We introduce
a set of normalized $2$-particle eigenstates of $X_f^{(2)}$ belonging to ${\cal H}_{{\frac{1}{2}(3+\varepsilon)}}^{(2)}$ as 
\bea\label{plusminus2p}
|n;0\pm\rangle_{[\varepsilon]}^{(2)}&=&\frac{1}{\sqrt 2}\cdot\big(
|n;010\rangle_{[\varepsilon]}^{(2)}\pm 
|n;001\rangle_{[\varepsilon]}^{(2)}\big),\nonumber\\
|n;1\pm\rangle_{[\varepsilon]}^{(2)}&=&\frac{1}{\sqrt 2}\cdot\big(
|n;110\rangle_{[\varepsilon]}^{(2)}\pm 
|n;101\rangle_{[\varepsilon]}^{(2)}\big),
\eea
while, a set of normalized  $3$-particle eigenstates of $X_f^{(3)}$ belonging to ${\cal H}_{{\frac{1}{2}(3+\varepsilon)}}^{(3)}$ is
\bea\label{plusminus3p}
|n;0\pm\rangle_{[\varepsilon]}^{(3)}&=&\frac{1}{\sqrt 2}\cdot\big(
|n;010\rangle_{[\varepsilon]}^{(3)}\pm 
|n;001\rangle_{[\varepsilon]}^{(3)}\big),\nonumber\\
|n;1\pm\rangle_{[\varepsilon]}^{(3)}&=&\frac{1}{\sqrt 2}\cdot\big(
|n;110\rangle_{[\varepsilon]}^{(3)}\pm 
|n;101\rangle_{[\varepsilon]}^{(3)}\big),\nonumber\\
|n;2\pm\rangle_{[\varepsilon]}^{(3)}&=&\frac{1}{\sqrt 2}\cdot\big(
|n;210\rangle_{[\varepsilon]}^{(3)}\pm 
|n;201\rangle_{[\varepsilon]}^{(3)}\big).
\eea
From now on we  use the $\varepsilon=\pm 1$ sign employed in (\ref{2psusyz2z2}) and (\ref{3psusyz2z2}) to distinguish supersymmetric versus ${\mathbb Z}_2\times{\mathbb Z}_2$-graded vectors.\par
The Hilbert space ${\cal H}_{{\frac{1}{2}(3+\varepsilon)}}^{(2)}$ is spanned, see (\ref{finitefull},\ref{2psusyz2z2}), for $n\in {\mathbb N}_0$ by the vectors given in (\ref{plusminus2p}) and the vectors $|n;000\rangle_{[\varepsilon]}^{(2)},~|n;100\rangle_{[\varepsilon]}^{(2)},
~|n;200\rangle_{[\varepsilon]}^{(2)},~ |n;011\rangle_{[\varepsilon]}^{(2)}$.\par
The Hilbert space ${\cal H}_{{\frac{1}{2}(3+\varepsilon)}}^{(3)}$ is spanned, see (\ref{finitefull},\ref{3psusyz2z2}), for $n\in {\mathbb N}_0$ by the vectors given in (\ref{plusminus3p}) and the vectors $|n;000\rangle_{[\varepsilon]}^{(3)},~|n;100\rangle_{[\varepsilon]}^{(3)},
~|n;200\rangle_{[\varepsilon]}^{(3)},~|n;300\rangle_{[\varepsilon]}^{(3)},~ |n;011\rangle_{[\varepsilon]}^{(3)}, ~|n;111\rangle_{[\varepsilon]}^{(3)}$.\par
Let's set aside, for the moment, the $2$-particle vectors $ |n;011\rangle_{[\varepsilon]}^{(2)}$ and the $3$-particle vectors $ |n;011\rangle_{[\varepsilon]}^{(3)},  ~|n;111\rangle_{[\varepsilon]}^{(3)}$. The remaining $M=2,3$-particle vectors are eigenvectors of the corresponding ($M=2,3$) $H_4^{(M)},N_f^{(M)},N_T^{(M)},X_f^{(M)}$ operators. Their eigenvalues are reported as entries in the tables below.\par
In the $2$-particle cases with $\varepsilon=\pm 1$ we have:
\bea\label{2psecondseteigen}&
\begin{array}{|c|c|c|c|c|}\hline  &H_4^{(2)}&N_f^{(2)}&N_T^{(2)}&X_f^{(2)}\\ \hline 
|n;000\rangle_{[\varepsilon]}^{(2)}:&n&0&0&0\\ \hline 
|n;100\rangle_{[\varepsilon]}^{(2)}:&n&0&2&0\\ \hline 
|n;200\rangle_{[\varepsilon]}^{(2)}:&n&0&4&0\\ \hline 
|n;0+\rangle_{[\varepsilon]}^{(2)}:&n+1&1&1&1\\ \hline 
|n;0-\rangle_{[\varepsilon]}^{(2)}:&n+1&1&1&-1\\ \hline 
|n;1+\rangle_{[\varepsilon]}^{(2)}:&n+1&1&3&1\\ \hline 
|n;1-\rangle_{[\varepsilon]}^{(2)}:&n+1&1&3&-1\\ \hline 
\end{array}&
\eea
In the $3$-particle cases with $\varepsilon=\pm 1$ we have:
\bea\label{3psecondseteigen}&
\begin{array}{|c|c|c|c|c|}\hline  &H_4^{(3)}&N_f^{(3)}&N_T^{(3)}&X_f^{(3)}\\ \hline 
|n;000\rangle_{[\varepsilon]}^{(3)}:&n&0&0&0\\ \hline 
|n;100\rangle_{[\varepsilon]}^{(3)}:&n&0&2&0\\ \hline 
|n;200\rangle_{[\varepsilon]}^{(3)}:&n&0&4&0\\ \hline 
|n;300\rangle_{[\varepsilon]}^{(3)}:&n&0&6&0\\ \hline 
|n;0+\rangle_{[\varepsilon]}^{(3)}:&n+1&1&1&1\\ \hline 
|n;0-\rangle_{[\varepsilon]}^{(3)}:&n+1&1&1&-1\\ \hline 
|n;1+\rangle_{[\varepsilon]}^{(3)}:&n+1&1&3&1\\ \hline 
|n;1-\rangle_{[\varepsilon]}^{(3)}:&n+1&1&3&-1\\ \hline 
|n;2+\rangle_{[\varepsilon]}^{(3)}:&n+1&1&5&1\\ \hline 
|n;2-\rangle_{[\varepsilon]}^{(3)}:&n+1&1&5&-1\\ \hline 
\end{array}&
\eea
Since the eigenvalues reported in the two tables (\ref{2psecondseteigen},\ref{3psecondseteigen}) do not depend on $\varepsilon$, we cannot distinguish
the supersymmetric with respect to the ${\mathbb Z}_2\times{\mathbb Z}_2$-graded Hilbert spaces with measurements performed on the above sets of states.\par
Let's now consider the $2$-particle  supersymmetric states $|n;011\rangle_{[-1]}^{(2)}$. They are eigenvectors of
$H_4^{(2)},~N_f^{(2)},~N_T^{(2)},~X_f^{(2)}$ with eigenvalues
\bea\label{2p011minus}&
\begin{array}{|c|c|c|c|c|}\hline  &H_4^{(2)}&N_f^{(2)}&N_T^{(2)}&X_f^{(2)}\\ \hline 
|n;011\rangle_{[-1]}^{(2)}:&n+2&2&2&0\\ \hline 
\end{array}&
\eea
The corresponding ${\mathbb Z}_2\times{\mathbb Z}_2$-graded $2$-particle states $|n;011\rangle_{[+1]}^{(2)}$ are eigenvectors of $H_4^{(2)},~N_f^{(2)},~N_T^{(2)}$ with eigenvalues
\bea\label{2p011plus}&
\begin{array}{|c|c|c|c|c|}\hline  &H_4^{(2)}&N_f^{(2)}&N_T^{(2)}&X_f^{(2)}\\ \hline 
|n;011\rangle_{[+1]}^{(2)}:&n+2&2&2&\times\\ \hline 
\end{array}&
\eea
The difference with the previous case lies on the $X_f^{(2)}$ operator. In the supersymmetric case, according to its 
$0$-grading, $X_f^{(2)}$ is an observable for the supersymmetric Hilbert space  ${\cal H}_1^{(2)}$. In the  ${\mathbb Z}_2\times{\mathbb Z}_2$-graded case $X_f^{(2)}$ is not defined as an operator acting on ${\cal H}_2^{(2)}$.\par
This can be understood as follows. It was shown in (\ref{2psusyz2z2}) that the vectors $|n;011\rangle_{[\varepsilon]}^{(2)}$ are proportional to
\bea
|n;011\rangle_{[-1]}^{(2)}~\propto~ (v_{12}-v_{15}),&&|n;011\rangle_{[+1]}^{(2)}~\propto~ (v_{12}+v_{15}).
\eea
The operator $X_f^{(2)}$, acting on the component vectors $v_{12},~v_{15}$ gives
\bea
&X_f^{(2)}v_{12}=X_f^{(2)}v_{15}=v_{11}+v_{16}.&
\eea
The relative $-$ sign entering $
|n;011\rangle_{[-1]}^{(2)}$ makes the two contributions vanish, leading to the $0$ eigenvalue shown in table (\ref{2p011minus}). On the other hand, the $+$ sign entering $
|n;011\rangle_{[+1]}^{(2)}$ implies that 
\bea
X_f^{(2)}|n;011\rangle_{[+1]}^{(2)}&\propto& (v_{11}+v_{16}).
\eea
By inspecting (\ref{2psusyz2z2}) it is clear that neither the vector $v_{11}$, nor $v_{16}$ belong to the ${\mathbb Z}_2\times {\mathbb Z}_2$-graded Hilbert space ${\cal H}_2^{(2)}$. Therefore,
\bea
X_f^{(2)}|n;011\rangle_{[+1]}^{(2)}&\not\in& {\cal H}_2^{(2)}.
\eea
Analogous formulas are obtained for the $3$-particle operators $ |n;011\rangle_{[\varepsilon]}^{(3)},  ~|n;111\rangle_{[\varepsilon]}^{(3)}$. In that case the eigenvalues are read from the table
\bea\label{3p}&
\begin{array}{|c|c|c|c|c|}\hline  &H_4^{(3)}&N_f^{(3)}&N_T^{(3)}&X_f^{(3)}\\ \hline 
|n;011\rangle_{[-1]}^{(3)}:&n+2&2&2&0\\ \hline 
|n;111\rangle_{[-1]}^{(3)}:&n+2&2&4&0\\ \hline \hline
|n;011\rangle_{[+1]}^{(3)}:&n+2&2&2&\times\\ \hline 
|n;111\rangle_{[+1]}^{(3)}:&n+2&2&4&\times\\ \hline 
\end{array}&
\eea
The action of $X_f^{(3)}$ on the vectors $|n;011\rangle_{[+1]}^{(3)}$ and $|n;111\rangle_{[+1]}^{(3)}$ is not defined in ${\cal H}_2^{(3)}$.

\subsection{Observables detecting ${\mathbb Z}_2$-graded versus ${\mathbb Z}_2\times{\mathbb Z}_2$-graded Hilbert spaces}

A legitimate observable, such as ${X}_f^{(2)}$, of the supersymmetric theory fails to be an operator of the ${\mathbb Z}_2\times{\mathbb Z}_2$-graded theory. We now show the existence of observables which, acting on both the supersymmetric and ${\mathbb Z}_2\times{\mathbb Z}_2$-graded Hilbert spaces, produce different, ${\varepsilon}$-dependent, eigenvalues on certain states of the models. The observables in questions, presented in (\ref{2psecondsetop2},\ref{3psecondsetop2}) as $W_{2f}, Z_{2f}, W_{3f}, Z_{3f}$ are recovered, together with $Y_{2f}, Y_{3f}$, from squaring the ${M=2,3}$ operators $X_f^{(M)}$. All these observables have $0$-grading in the supersymmetric case and, see (\ref{0011gradingobs}), $00$-grading in the ${\mathbb Z}_2\times{\mathbb Z}_2$-graded theory. The $|n;r_0r_1r_2\rangle_{[\varepsilon]}^{(M=2,3)}$ vectors spanning the Hilbert spaces ${\cal H}_{\frac{1}{2}(3+\varepsilon)}^{M=2,3}$ are eigenvectors of $Y_{2f},W_{2f},Z_{2f}$ (for $M=2$) and $Y_{3f},W_{3f},Z_{3f}$ (for $M=3$). Their eigenvalues are given as entries in the tables below. \par
For the $2$-particle case we have
\bea\label{2psecondsetywz}&
\begin{array}{|c|c|c|c|c|c|}\hline  &Y_{2f}&{ W}_{2f}&Z_{2f}\\ \hline 
|n;000\rangle_{[\varepsilon]}^{(2)}:&0&0&0\\ \hline 
|n;100\rangle_{[\varepsilon]}^{(2)}:&0&0&0\\ \hline 
|n;200\rangle_{[\varepsilon]}^{(2)}:&0&0&0\\ \hline 
|n;010\rangle_{[\varepsilon]}^{(2)}:&1&0&1\\ \hline 
|n;001\rangle_{[\varepsilon]}^{(2)}:&1&0&1\\ \hline 
|n;110\rangle_{[\varepsilon]}^{(2)}:&1&0&1\\ \hline 
|n;101\rangle_{[\varepsilon]}^{(2)}:&1&0&1\\ \hline 
|n;011\rangle_{[\varepsilon]}^{(2)}:&2&2\varepsilon&2+2\varepsilon\\ \hline 
\end{array}&
\eea\\
For the $3$-particle case we have
\bea\label{3psecondsetywz}&
\begin{array}{|c|c|c|c|c|c|}\hline  &Y_{3f}&W_{3f}&Z_{3f}\\ \hline 
|n;000\rangle_{[\varepsilon]}^{(3)}:&0&0&0\\ \hline 
|n;100\rangle_{[\varepsilon]}^{(3)}:&0&0&0\\ \hline 
|n;200\rangle_{[\varepsilon]}^{(3)}:&0&0&0\\ \hline 
|n;300\rangle_{[\varepsilon]}^{(3)}:&0&0&0\\ \hline 
|n;010\rangle_{[\varepsilon]}^{(3)}:&1&0&1\\ \hline 
|n;001\rangle_{[\varepsilon]}^{(3)}:&1&0&1\\ \hline 
|n;110\rangle_{[\varepsilon]}^{(3)}:&1&0&1\\ \hline 
|n;101\rangle_{[\varepsilon]}^{(3)}:&1&0&1\\ \hline 
|n;210\rangle_{[\varepsilon]}^{(3)}:&1&0&1\\ \hline 
|n;201\rangle_{[\varepsilon]}^{(3)}:&1&0&1\\ \hline 
|n;011\rangle_{[\varepsilon]}^{(3)}:&2&2\varepsilon&2+2\varepsilon\\ \hline 
|n;111\rangle_{[\varepsilon]}^{(3)}:&2&2\varepsilon&2+2\varepsilon\\ \hline 
\end{array}&
\eea\\
As a consequence of (\ref{2psecondsetywz}) the following theoretical scenario can be applied. 
At a given $\varepsilon=\pm 1$ the $2$-particle state $|n;011\rangle_{[\varepsilon]}^{(2)}$
can be selected as the unique state possessing, see (\ref{n4multiparteigen}), the set of ($n+2,2,1,1$) eigenvalues for,
respectively, the operators ($H_4^{(2)},N_f^{(2)}, N_v^{(2)},N_w^{(2)}$).  \par
A system under investigation can be prepared at first in this eigenstate. Once this is accomplished, it is possible to determine whether  the $2$-particle system so prepared is composed by ordinary  bosons/fermions (supersymmetric case) or by ${\mathbb Z}_2\times{\mathbb Z}_2$-graded particles by performing a measurement of  the $W_{2f}$ 
(or, equivalently, $Z_{2f}$) observable. The output of the measurement can be expressed in terms of the $\varepsilon$ sign ($-1$ for supersymmetry, $+1$ otherwise). \par
The term ``chirality" can be conveniently employed  to convey the difference between the two cases. One can then
say that the supersymmetric versus the ${\mathbb Z}_2\times{\mathbb Z}_2$-graded variants of the model have opposite chirality.\par
The same scenario works for $3$-particle systems prepared in the $|n;011\rangle_{[\varepsilon]}^{(3)}$ and
$|n;111\rangle_{[\varepsilon]}^{(3)}$ eigenstates.

\section{Conclusions}

The ${\mathbb Z}_2\times{\mathbb Z}_2$-graded parastatistics requires the particles to be accommodated into the 
${\mathbb Z}_2\times{\mathbb Z}_2$-graded setting presented in Appendix {\bf A}.  The particles are $00$-graded ordinary bosons, $11$-graded exotic  bosons and two types of fermions ($10$- and $01$-graded, with fermions of different types which mutually commute). The presence of some given  ${\mathbb Z}_2\times{\mathbb Z}_2$-graded invariance is a sufficient, but not necessary condition, for the application of a  ${\mathbb Z}_2\times{\mathbb Z}_2$-graded parastatistics.\par
We tested the consequences of the ${\mathbb Z}_2\times{\mathbb Z}_2$-graded parastatistics by analizing the simple model of a $4\times 4$ matrix quantum oscillator which, in its single-particle sector, admits three different interpretations: as a bosonic, as a ${\cal N}=4$ supersymmetric and as a ${\mathbb Z}_2\times{\mathbb Z}_2$-graded one-dimensional Poincar\'e-invariant theory.  These three interpretations give equivalent, physically indistinguishable, single-particle theories. The simplicity of the model allows to construct
non-interacting multiparticle sectors by applying Hopf algebra coproducts. The multiparticle states are defined in terms of braided tensor products; for each one of the three above cases the braiding is related with the respective grading (bosonic, supersymmetric, ${\mathbb Z}_2\times{\mathbb Z}_2$). As a consequence, each grading produces a different multiparticle theory.\par
The bosonic grading is easily distinguished from the other two cases since it produces different degeneracies of the energy levels;  this is easily understood by noting, in the bosonic theory,  the existence of extra states which are not allowed, due to Pauli's exclusion principle, when fermions are present.\par
The difference between the supersymmetric and ${\mathbb Z}_2\times{\mathbb Z}_2$-grading is much subtler and more elusive. The energy levels and their degeneracies coincide in both cases. Despite of that it is still possible to construct certain types of observables such as $X_f^{(2)}$  in the discussion of Section {\bf 6}. These observables act on both the supersymmetric and 
${\mathbb Z}_2\times{\mathbb Z}_2$-graded multiparticle Hilbert spaces and, on certain states, produce different outputs (eigenvalues)  which depend on the grading. A measurement of these observables allows to determine whether the system under investigation is composed by ordinary bosons/fermions or by ${\mathbb Z}_2\times{\mathbb Z}_2$-graded particles.\par
We also pointed out in Section {\bf 6} that the ${\mathbb Z}_2\times{\mathbb Z}_2$-grading forces the observables of the theory to be superselected and belonging to the $00$-graded sector.\par
In principle the construction based on the coproduct can offer the guidelines to correctly (anti)symmetrize the ${\mathbb Z}_2\times{\mathbb Z}_2$-graded multiparticle states in more complicated Hamiltonians such as
those, obtained in \cite{aktquant}, which present ${\mathbb Z}_2\times{\mathbb Z}_2$-graded invariant multiparticle interacting terms. This extension will be left for a future work.\par
The elusiveness of the ${\mathbb Z}_2\times{\mathbb Z}_2$-graded parastatistics (the observables detecting it should be carefully cherry picked) implies that it can be easily get unnoticed even if present. Fermions, in Nature, can be either fundamental, as those entering the Standard Model, or emergent, e.g. as collective modes in materials.\par
Concerning fundamental spinors, in the light of the results here presented,  the spin-statistics connection for relativistic ${\mathbb Z}_2\times{\mathbb Z}_2$-graded Quantum Field Theories would demand a careful (re)evaluation. If the observables of the theories are requested to obey a superselection rule and possess the $00$-grading (just like their nonrelativistic counterparts), it looks possible to realize consistent relativistic ${\mathbb Z}_2\times{\mathbb Z}_2$-graded models
which take into account the commuting properties of the fermions belonging to different gradings ($01$- and $10$-). No matter which is their grading, these fermions would continue to be half-integer spin particles. It seems that the simplest setting to check the consistency of this scheme should be formulated for a ${\mathbb Z}_2\times{\mathbb Z}_2$-graded extension of the Wess-Zumino model \cite{wz}. Investigations are currently underway.\par
Concerning emergent spinors, a connection could be made with the \cite{kit} proposal  to use Majorana fermions for the construction of a Topological Quantum Computer which offers topological protection from decoherence. The connection with braiding properties and statistics is elucidated in \cite{kau}. This could be a playground for
${\mathbb Z}_2\times {\mathbb Z}_2$-graded parastatistics and possibly general ${\mathbb Z}_2^n$-graded parastatistics
with $n\geq 3$. Currently ${\mathbb Z}_2^n$-graded structures are under intense investigation,
see \cite{{bru},{AAD},{AAD2},{brgr}}. At the moment the formulation, for a generic $n$,
of a ${\mathbb Z}_2^n$-graded mechanics following the lines of \cite{aktclass,aktquant} for $n=2$ is still  lacking.  
\par

\par

~\par
~\par
\par {\Large{\bf Acknowledgments}}
{}~\par{}~\par

The author is grateful for constant discussions with Naruhiko Aizawa and Zhanna Kuznetsova.\par
This investigation was also partly inspired by a question raised by the next-to-my-office-room colleague Itzhak Roditi when we were not yet quarantined due to COVID 19.\par
This work is supported by CNPq (PQ grant 308095/2017-0). 

\par

~\par
~\par

  \renewcommand{\theequation}{A.\arabic{equation}}
  \setcounter{equation}{0}  

\textcolor{black}{
{\Large{\bf{Appendix A: Graded Lie superalgebras}}} }\par
~\par

This paper deals with the physical differences induced, in the multiparticle sector of a quantum theory, of three different graded Lie (super)algebras. It makes sense to discuss them in a unified framework to pinpoint, as much as possible, their common properties.\par
The graded structures under consideration will be denoted as ${\mathbb Z}_2^{p}$ for $p=0,1,2$: 
\bea\label{grp}
&{\mathbb Z}_2^0={\bf 1},\qquad {\mathbb Z}_2^1={\mathbb Z}_2,\qquad
{\mathbb Z}_2^2={\mathbb Z}_2\times {\mathbb Z}_2.&
\eea
They respectively correspond to\\
~\\
{\it $~~$  i})  for $p=0$, ordinary Lie algebras,\\
{\it $~$ ii})  for $p=1$, ordinary Lie superalgebras \cite{kac} and,\\
{\it iii})  for $p=2$, the so-called, see \cite{{rw1},{rw2}}, ${\mathbb Z}_2\times {\mathbb Z}_2$-graded Lie superalgebras. \\
~\\
A graded Lie superalgebra ${\mathfrak g}^{[p]}$ denotes a Lie algebra of graded brackets (represented by (anti)commutators) satisfying the  respective graded Jacobi identities for $p=0,1,2$.\par
The $p=1,2$ graded superalgebras are decomposed in the graded sectors
\bea\label{gradingdecomposition}
{\mathfrak g}^{[1]} &=& {\mathfrak g}^{[1]}_{[0]}\oplus {\mathfrak g}^{[1]}_{[1]},\nonumber\\
{\mathfrak g}^{[2]} &=& {\mathfrak g}^{[2]}_{[00]}\oplus {\mathfrak g}^{[2]}_{[11]}\oplus  {\mathfrak g}^{[2]}_{[10]}\oplus {\mathfrak g}^{[2]}_{[01]}.
\eea
Formally one can also set ${\mathfrak g}^{[0]}={\mathfrak g}^{[0]}_{[0]}$.\par
A generator $g\in {\mathfrak{g}}^{[p]}$ is associated with its ${\epsilon}^{[p]}_g$ grading which takes values
\bea
&p=0:~{\epsilon}^{[0]}_g = \{0\}; \qquad \quad p=1:~{ \epsilon}^{[1]}_g=\{ 0, 1\};\qquad \quad p=2:~ {\epsilon}^{[2]}_g =\{ 00,11,10,01\}.&
\eea
The operations on the $\epsilon^{[p]}_g$'s gradings are the sum  and the inner scalar product. Let us set\\
~\\
 $~${\it $~$i}) ~~for~~ $p=1$: \quad
${\epsilon}^{[1]}_A =\alpha,~\quad {\epsilon}^{[1]}_B=\beta\quad $ with $\quad \alpha,\beta=0,1\quad $ and\\
~\\
{\it ii})~~ for~~ $p=2$: \quad $ {\epsilon}^{[2]}_A =(\alpha_1,\alpha_2)$,  ~~$ {\epsilon}^{[2]}_B =(\beta_1,\beta_2)\quad $
with $\quad \alpha_1,\alpha_2,\beta_1,\beta_2=0,1$. 
\\~\par
We have
\bea
{\epsilon}^{[1]}_A+{\epsilon}^{[1]}_B=\alpha+\beta\quad \qquad\qquad\quad~({\textrm{mod}}~ 2), \qquad && {\epsilon}^{[1]}_A\cdot {\epsilon}^{[1]}_B=\alpha\beta,\nonumber\\
{\epsilon}^{[2]}_A+{\epsilon}^{[2]}_B=(\alpha_1+\beta_1,\alpha_2+\beta_2) \quad({\textrm{mod}}~ 2), \qquad && {\epsilon}^{[2]}_A\cdot {\epsilon}^{[2]}_B=\alpha_1\beta_1+\alpha_2\beta_2.
\eea

Let $A,B,C\in  {\mathfrak{g}}^{[p]}$ be three generators of a graded Lie superalgebra, with respective gradings
$\epsilon_A,\epsilon_B,\epsilon_C$ (for simplicity the now unecessary superscript ``$[p]$" is dropped). The brackets\bea
[\cdot,\cdot\} &:&  {\mathfrak{g}}^{[p]}\times  {\mathfrak{g}}^{[p]}\rightarrow  {\mathfrak{g}}^{[p]} 
\eea
which define the graded Lie superalgebra must be compatible with the ${\mathbb Z}_2^{p}$ grading. This requires the 
following properties to be satisfied: \\
{\it i}) (anti)symmetry
\bea\label{mixedbracket}
[A,B\}&:=& AB -(-1)^{\epsilon_A\cdot\epsilon_B}BA,
\eea
so that
\bea
\relax [B,A\} &=& (-1)^{\epsilon_A\cdot\epsilon_B+1}[A,B\};
\eea
the grading of the $[A,B\}$ generator is
\bea\label{gradingsum}
\epsilon_{[A,B\}} &=& \epsilon_A+\epsilon_B.
\eea
{\it ii}) graded Jacobi identities given by
\bea
\label{gradedjacobi}
 (-1)^{\epsilon_C\cdot\epsilon_A}[A,[B,C\}\}+
 (-1)^{\epsilon_A\cdot\epsilon_B}[B,[C,A\}\}+
 (-1)^{\epsilon_B\cdot\epsilon_C}[C,[A,B\}\}&=&0.
\eea

A graded vector space $V^{[p]}$ is a representation space for the graded Lie (super)algebra ${\mathfrak g}^{[p]}$ if
for any pair of generators $A,B\in {\mathfrak{g}}^{[p]}$ the bracket (\ref{mixedbracket}) is realized by (anti)commutators of the operators ${\widehat A},{\widehat B}$ through the mapping
\bea
&A\mapsto {\widehat A},\quad B\mapsto {\widehat B},\qquad {\textrm{with}}\quad {\widehat A},{\widehat B}: V^{[p]}\rightarrow V^{[p]}.
\eea
The grading of  $V^{[p]}$ is compatible with the grading of ${\mathfrak{g}}^{[p]}$. This implies that
for any pair of vectors $v, w\in V^{[p]}$  with respective gradings $\epsilon_v,\epsilon_{w}$ and for any operator ${\widehat A}$ of grading $\epsilon_A$ one has
\bea\label{algvec}
w={\widehat A}v &\Rightarrow &\epsilon_{w} = \epsilon_A+\epsilon_v.
\eea
For $p=1,2$ the graded vector spaces are decomposed as
\bea\label{gradedvector}
{\mathbb Z}_2:\quad V^{[1]}=V^{[1]}_{[0]}\oplus V^{[1]}_{[1]},\qquad && {\mathbb Z}_2\times{\mathbb Z}_2:\quad 
{V}^{[2]} = {V}^{[2]}_{[00]}\oplus {V}^{[2]}_{[11]}\oplus {V}^{[2]}_{[10]}\oplus {V}^{[2]}_{[01]}.
\eea
For $p=2$ the graded (anti)commutators defined on the $A,B$ generators of a ${\mathbb Z}_2\times{\mathbb Z}_2$-graded Lie superalgebra can be read from the table
\bea
&\begin{array}{|c|c|c|c|c|}\hline 
 A\backslash B&[00]&[10]&[01]&[11] \\  \hline
[00]&[\cdot,\cdot]&[\cdot,\cdot]&[\cdot,\cdot]&[\cdot,\cdot]\\ \hline
[10]&[\cdot,\cdot]&\{\cdot,\cdot\}&[\cdot,\cdot]&\{\cdot,\cdot\} \\  \hline
[01]&[\cdot,\cdot]&[\cdot,\cdot]&\{\cdot,\cdot\}&\{\cdot,\cdot\} \\  \hline
[11]&[\cdot,\cdot]&\{\cdot,\cdot\}&\{\cdot,\cdot\}&[\cdot,\cdot] \\ \hline
\end{array}&
\eea\par
~
\par
~\par

  \renewcommand{\theequation}{B.\arabic{equation}}
  \setcounter{equation}{0}  

\textcolor{black}{
{\Large{\bf{Appendix B:  a summary of Hopf algebras and braided tensors}}} }\par
This paper relies on coproducts for the construction of  multiparticle states of a quantum system and the hermitian operators acting on them. The coproduct is a costructure entering the definition of Hopf algebra. To make the paper self-consistent the notion of Hopf algebra is here briefly recalled (for a more comprehensive treatment, see \cite{{swe},{abe}}). \par
The Hopf algebras under consideration in this work are the Universal Enveloping Algebras ${\cal U}({\mathfrak g}^{[p]})$  of a ${\mathbb Z}_2^p$-graded (with $p=0,1,2$), see (\ref{grp}), Lie superalgebra.\par
A Universal Enveloping Lie Algebra ${\cal U}({\mathfrak{g}^{[p]}})$ over a field which, for our purposes, is assumed to be ${\mathbb C}$, is a unital associative algebra containing the identity ${\bf 1}$ and the polynomials of the ${\mathfrak{g}^{[p]}}$ generators modulo their (anti)commutation relations. In the following, and throughout the text, a generic element of ${\cal U}({\mathfrak{g}^{[p]}})$ is denoted as ``$U$", while the symbol ``$g$" is reserved to the generators of ${\mathfrak{g}^{[p]}}$.\par
As a Hopf algebra ${\cal U}\equiv {\cal U}({\mathfrak{g}^{[p]}})$ possesses:\\
- two algebraic structures, the associative multiplication $m$ and the unit $\eta$, where
\bea
m:{\cal U}\otimes {\cal U}\rightarrow {\cal U}, \quad (m: U_A\otimes U_B\mapsto U_A\cdot U_B),&\quad&
\eta: {\cal U}\rightarrow {\mathbb C}\quad (\eta: {\bf 1}\mapsto 1),
\eea
- two algebraic costructures, the coproduct $\Delta$ and the counit $\varepsilon$
\bea\label{costructures}
\Delta:{\cal U}\rightarrow {\cal U} \otimes  {\cal U}, \quad &
\varepsilon: {\mathbb C}\rightarrow {\cal U},
\eea
- an operation, the antipode $S$, relating structures and costructures,
\bea
S: {\cal U}\rightarrow {\cal U}.
\eea
The compatibility of structures and costructures is guaranteed by a set of properties relating them. We have
\bea&\label{deltauu}
   \Delta(U_AU_B)=\Delta(U_A)\Delta(U_B),\qquad
  \varepsilon(U_AU_B)=\varepsilon(U_A)\varepsilon(U_B),\qquad
   S(U_AU_B)= S(U_B)S(U_A)&\nonumber\\&&
\eea
and
\bea\label{coassociativity}
   (\Delta\otimes id)\Delta(U)&=&(id\otimes \Delta)\Delta(U) \qquad \qquad \quad,\nonumber\\
   (\varepsilon\otimes id)\Delta(U)&=&(id\otimes\varepsilon)\Delta(U)=U \nonumber\\
   m(S\otimes id)\Delta(U)&=&m(id\otimes
   S)\Delta(U)=\varepsilon(U)\bf{1}.
\eea
The first relation is the coassociativity of the coproduct.

The action on the identity ${\bf 1}$ is
\bea\label{deltaid}
   &\Delta({\bf 1})={\bf 1}\otimes{\bf 1}, \qquad\qquad\quad~~~~
  \varepsilon({\bf 1})=1,\qquad\qquad
   S({\bf 1})={\bf 1}.&
\eea

The action on a primitive element, i.e. a generator $g\in {\mathfrak{g}^{[p]}}$, is
\bea\label{deltag}
   &\Delta({ g})={\bf 1}\otimes{g}+g\otimes {\bf 1}, \qquad\qquad
  \varepsilon({g})=0,\qquad\qquad
   S({g})={-g}.&
\eea
Following \cite{maj}, to properly (anti)symmetrize bosons and fermions in the language of the coproduct, the notion of braided tensor (which naturally incorporates a braid statistics) has to be used. The ${\mathbb Z}_2$-grading is the simplest non-trivial example of braiding; the ${\mathbb Z}_2\times {\mathbb Z}_2$-grading is the next simplest case.
In a braided tensor product,
\bea
(U_A\otimes U_B)(U_C\otimes U_D) &=& U_A\Psi(U_B\otimes U_C)U_D,
\eea
$U_B$ and $U_C$ are braided by an operator $\Psi$ acting on their tensor product; $\Psi$ is called the ``braiding operator" (see \cite{maj} for details).\par
In our applications to the ${\mathbb Z}_2^{[p]}$, $p=1,2$ gradings, the braiding simply reads
\bea\label{epsilonbraiding}
(U_A\otimes U_B)(U_C\otimes U_D) &=& (-1)^{\epsilon_B\cdot\epsilon_C} (U_A U_C)\otimes (U_BU_D)
\eea
and corresponds to a sign.\par
For the creation operator $f^\dagger$ introduced in (\ref{bbff}), with $(f^\dagger)^2=0$, in the bosonic interpretation
(for $\epsilon_{f^\dagger}=0$) the coproduct gives
\bea\label{bosonicfdagger}
\Delta (({f^\dagger})^2)&=& ({\bf 1}\otimes{f^\dagger}+f^\dagger\otimes {\bf 1}) ({\bf 1}\otimes{f^\dagger}+f^\dagger\otimes {\bf 1})=2f^\dagger\otimes f^\dagger\neq 0.
\eea
In the fermionic interpretation
(for $\epsilon_{f^\dagger}=1$) the coproduct gives
\bea\label{fermionicfdagger}
\Delta (({f^\dagger})^2)&=& ({\bf 1}\otimes{f^\dagger}+f^\dagger\otimes {\bf 1}) ({\bf 1}\otimes{f^\dagger}+f^\dagger\otimes {\bf 1})=f^\dagger\otimes {\bf 1}\cdot {\bf 1}\otimes{f^\dagger}+{\bf 1}\otimes{f^\dagger}\cdot f^\dagger\otimes {\bf 1}=\nonumber\\
&&f^\dagger\otimes f^\dagger-f^\dagger\otimes f^\dagger=0.
\eea
The physical consequence is that the coproduct, combined with the braided tensor, encodes the Pauli exclusion principle for fermions.\par
The permutations of spaces for the tensor products ${\cal U} \otimes \ldots \otimes  {\cal U}$ of a graded Universal Enveloping Lie superalgebra ${\cal U}({\mathfrak{g}}^{[p]})$ are defined as follows.\par
In the case of the product of two tensors the permutation operator is
\bea\label{S12}
S_{12}^{(2)}&:& U_A\otimes U_B \mapsto (-1)^{\epsilon_A\cdot\epsilon_B}U_B\otimes U_A, \quad (U_{A,B}\in {\cal U}\quad{\textrm{and}}\quad S_{12}^{(2)}\cdot S_{12}^{(2)}={\bf 1} ).
\eea
This action extends linearly to linear combinations of elements of ${\cal U}\otimes {\cal U}$. One should note
in the definition the presence of the $\pm 1$ sign induced by the respective $\epsilon_{A,B}$ gradings of  $U_{A,B}$.
\par
In the case of the product of three tensors we have two generators, $S_{12}^{(3)}$ and $ S_{23}^{(3)}$, given by
\bea\label{S12S13}
S_{12}^{(3)}&:& U_A\otimes U_B \otimes U_C\mapsto (-1)^{\epsilon_A\cdot\epsilon_B}U_B\otimes U_A\otimes U_C,\nonumber\\
S_{23}^{(3)}&:& U_A\otimes U_B \otimes U_C\mapsto (-1)^{\epsilon_B\cdot\epsilon_C}U_A\otimes U_C\otimes U_B.
\eea
The $6$ elements of the ${\bf S}_3$ permutation group of three letters are conveniently expressed in terms of the operators $S,T$:
\bea
&S:=S_{12}^{(3)}, \quad T:=S_{12}^{(3)}\cdot S_{23}^{(3)}, \quad {\textrm{so that}} \quad S^2=T^3={\bf 1}\quad {\textrm{and}}\quad
\{{\bf 1},S, T, ST, T^2, ST^2\} \in {\bf S}_3. &\nonumber\\
&&
\eea
The extension to permutations of tensor products with $M>3$ spaces is immediate.\par
Throughout the text, if the abstract Universal Enveloping Algebra ${\cal U}$ is represented on a vector space $V$
under the $R$ representation, a hat denotes the action (on the tensor products of $V$) of the operators induced by the coproduct:
\bea
&{\textrm{for}}\quad R: {\cal U}\rightarrow V,\qquad {\widehat \Delta}:= \Delta|_R\in End (V\otimes V),\qquad 
 {\textrm{with}}\quad  {\widehat{ \Delta(U)}}\in V\otimes V. &
\eea

\end{document}